\documentclass[aps,nofootinbib,notitlepage]{revtex4}
\usepackage[english]{babel}
\usepackage{amstext,amsmath,amssymb,amsfonts,bbm}
\usepackage[latin1]{inputenc}
\usepackage{graphicx}
\usepackage{epstopdf}
\usepackage{amsthm}
\usepackage{tocvsec2}
\usepackage{enumerate}
  \usepackage{dutchcal}
\usepackage{color}
\usepackage{xcolor}
\usepackage{mathtools}
\usepackage{bbm}
\usepackage{braket}
\usepackage{tcolorbox}
\usepackage{natbib}
\usepackage{graphicx}
\definecolor{darkgreen}{rgb}{0.0, 0.2, 0.13}
\definecolor{darkred}{rgb}{0.55, 0.0, 0.0}
\definecolor{carrotorange}{rgb}{0.93, 0.57, 0.13}
\usepackage{hyperref}
\hypersetup{
	colorlinks=true,
	linkcolor=blue,
	filecolor=magenta,
	urlcolor=cyan,
}
\usepackage[squaren,Gray]{SIunits}

\newcommand{\be}{\nopagebreak[3]\begin{equation}}
\newcommand{\ee}{\end{equation}}

\newcommand{\bee}{\nopagebreak[3]\begin{equation*}}
\newcommand{\eee}{\end{equation*}}

\newcommand{\ba}{\nopagebreak[3]\begin{eqnarray}}
\newcommand{\ea}{\end{eqnarray}}
\DeclareFontFamily{U}{rsfs}{}         
\DeclareFontShape{U}{rsfs}{m}{n}{<5> rsfs5 <6><7> rsfs7          %
  <8><9><10><10.95><12><14.4><17.28><20.74><24.88> rsfs10}{}     %
\DeclareMathAlphabet{\mathfs}{U}{rsfs}{m}{n}                     %
\newcommand{\mfs}[1]{\mathfs {#1}}                               %
\newcommand{\n}{{\nonumber}}

\newcommand{\sH}{{\mfs H}}

\newcommand{\sN}{{\mfs N}}

\newcommand{\sO}{{\mfs O}}

 \usepackage{braket}
\newcommand{\N}{\mathbb{N}}

\newcommand{\R}{\mathbb{R}}
\newcommand{\Z}{\mathbb{Z}}\def\be#1\ee{\begin{align}#1\end{align}}
\def\ba{\begin{eqnarray}}
\def\ea{\end{eqnarray}}

\def\lpr{\left(}
\def\rpr{\right)}

\def\R{\mathbb{R}}

\def\n{\nonumber}

\renewenvironment{thebibliography}[1]
         {\section*{References}\frenchspacing\small
          \begin{list}{[\arabic{enumi}]}
         {\usecounter{enumi}\parsep=2pt\topsep 0pt
         \settowidth{\labelwidth}{[#1]}
         \leftmargin=\labelwidth\advance\leftmargin\labelsep
         \rightmargin=0pt\itemsep=1pt\sloppy}}{\end{list}}

\numberwithin{equation}{section}

\newtheorem{coloro}{Corolary}

\newtheorem{prop}{Remark}

\begin{document}

\title{The landscape of polymer quantum cosmology}

\author{Lautaro Amadei}
\affiliation{{Aix Marseille Univ, Universit\'e de Toulon, CNRS, CPT, Marseille, France}}

\author{Alejandro Perez}
\affiliation{{Aix Marseille Univ, Universit\'e de Toulon, CNRS, CPT, Marseille, France}}

\author{Salvatore Ribisi}
\affiliation{{Aix Marseille Univ, Universit\'e de Toulon, CNRS, CPT, Marseille, France}}

\date{\today}

\begin{abstract}
   We show that the quantization ambiguities of loop quantum cosmology, when considered in wider generality, can be used
   to produce discretionary dynamical behaviour. There is an infinite dimensional space of ambiguities which parallels the 
   infinite list of higher curvature corrections in perturbative quantum gravity. There is however an ensemble of qualitative consequences which are generic in the sense that they are independent of  
   these ambiguities. Among these, one has well defined fundamental dynamics across the big bang, and the existence of extra microscopic quantum degrees of freedom that might be relevant in discussions about unitarity in quantum gravity. We show that (in addition to the well known bouncing solutions of the effective equations)  there are other generic type of solutions for sufficiently soft initial conditions in the matter sector (tunnelling solutions) where the scale factor goes through zero and the spacetime orientation is inverted. We also show that generically, a contracting semiclassical universe branches off at the big bang into a quantum superposition of universes with different quantum numbers.  
 Despite their lack of quantitative predictive power these models offer a fertile playground for the discussion of qualitative and conceptual issues in quantum gravity.
\end{abstract}


\maketitle


\section{Introduction}

Quantum field theories with local degrees of freedom generically suffer of divergences due to uncontrolled UV contributions to amplitudes. At the mathematical level the latter can be traced to the fact that interactions involve products of fields (operator valued distributions in quantum field theory) at the same spacetime point and that such products are ill-defined if constructed naively. The standard procedure of renormalization eliminates infinities from the physical amplitudes at the price of introducing counter terms with free parameters to be fixed by a series of renormalization conditions taken from physical inputs. In certain simple situations one can instead take due care in the definition of products of operator valued distributions and thus completely avoid from the very beginning UV divergences (see for instance \cite{scharf2014finite}). However, such procedure is not unique and free parameters arise too in the regularization procedure. These parameters must be fixed (in order to produce physical amplitudes) by the same number of renormalization conditions of standard textbook treatments.  In this way there is  a formal link between the number of counter-terms necessary to eliminate UV divergences and regularization ambiguities.

A key difficulty of canonical approaches to quantum gravity is that such intrinsic ambiguity of standard quantization recipes become in appearance  out of control. This is associated to the non-renormalizability of the gravitational interaction, which in the case of general relativity in metric variables, is illustrated by the fact that the family of general covariant functionals of $g_{ab}$ representing a possible action principle describing the quantum effective action is infinite dimensional and that the parameter controlling the dimensionality of such free couplings is the quantum gravity coupling itself. Namely,
\be\label{action1}
S[g_{ab}]=\frac{1}{2 \kappa} \int \sqrt{|g|} \left(R+\Lambda+\alpha_1 \ell_p^2 R^2+\alpha_2 \ell_p^4 R^3+\cdots+\beta_1\ell_p^2 R_{\mu\nu\alpha\sigma} R^{\mu\nu\alpha\sigma}\cdots\right) dx^4,
\ee 
where  only some representative terms have been written with dimensionless couplings  $\alpha_1, \alpha_2,\cdots$, $\beta_1, \beta_2, \cdots$, etc. Generic radiative corrections produce divergences that need to be cured by counter terms in correspondence with the infinite number of elements in the previous general action,  requiring infinitely many renormalization conditions, and compromising the predictive power of the approach\footnote{It is possible, however, that these couplings would flow under the renormalization group in a non trivial way toward some asymptotic fixed point characterized by a finite amount of parameters. Such perspective,  known as the {\em asymptotic safety} scenario \cite{Weinberg:1980gg},  is the subject of active investigations \cite{Niedermaier:2006wt}. }. However, from the previous formal discussion, an equivalent danger menaces non-perturbative formulations where UV divergences are avoided via 
clever  choices of variables and or mathematical structures, as the danger metamorphoses into that of ambiguities.    

As a complementary remark one must keep in mind that the previous analysis sometimes strongly depends on the `fundamental variables' chosen for quantization.  An example of this is the emblematic case of gravity in three dimensions where a naive metric variable analysis would have led to similar conclusions as in four dimensions. However, when the most general  action is written in terms of first order variables one discovers that there is only a finite dimensional set of possibilities. Namely
\be
S[e,\omega]=\frac{1}{2\kappa} \int e_I\wedge F_{IJ}(\omega) \epsilon^{IJK}+\Lambda \,e_I\wedge e_J \wedge e_K\epsilon^{IJK}+\alpha S_{\rm CS} (\omega),  
\ee
where $e^I_a$ is a triad field, $\omega_a^{IJ}$ is a Lorentz connection, and $S_{\rm CS}(\omega)$ is the Chern-Simons action. The theory is indeed integrable, has only  global or topological degrees of freedom, and its quantization is free of (infinite dimensional) ambiguities \cite{Witten:1988hc, Witten:1988hf}.
Strikingly, a similar finite dimensionality of the space of gravity actions is valid in first order variables in four dimensions where one has that the most general gravitational action is given by   
\ba\label{actiong}
S[e_a^A,\omega_a^{AB}]&=&\frac{1}{2 \kappa} \int \overbrace{ \epsilon_{IJKL} e^I\wedge e^J\wedge F^{KL}(\omega)}^{\rm Einstein}+\overbrace{{\Lambda} \, \epsilon_{IJKL} e^I\wedge e^J\wedge e^K\wedge e^L }^{\rm Cosmological\, Constant}+\overbrace{\alpha_1 \ e_I\wedge e_J\wedge F^{IJ}(\omega)}^{\rm Holst} \\\n  &+&\underbrace{\alpha_2 \ (d_{\omega} e^I \wedge d_{\omega} e_I \,- \ e_I\wedge e_J\wedge F^{IJ}(\omega))}_{\rm Nieh-Yan}  + \underbrace{\alpha_3 \ell_p^2 \ F(\omega)_{IJ} \wedge F^{IJ}(\omega)}_{\rm Pontrjagin}  +\underbrace{\alpha_4 \ell_p^2\ \epsilon_{IJKL} F(\omega)^ {IJ} \wedge F^{KL}(\omega)}_{\rm Euler}, 
\ea 
where $d_\omega e^I$ is the covariant exterior derivative of $e^I$ and $\alpha_1\cdots \alpha_4$ are dimensionless coupling constants.  For non-degenerate
tetrads Einstein's field equations follow from the previous action independently of the values of the $\alpha$'s:
the additional terms are called topological invariants describing global properties of the field configurations in spacetime. The $\alpha_1$-term is called the Holst term \cite{Holst:1995pc}, the $\alpha_2$-term is the Nieh-Yan invariant, the $\alpha_3$-term is the Pontryagin invariant, and the $\alpha_4$-term is the Euler invariant.  Inspite of not changing the equation of motion these terms can actually be interpreted as producing  canonical transformations in the phase space of gravity \footnote{  
In such a context the so-called Immirzi parameter \cite{Immirzi:1996di} corresponds to the combination $\gamma\equiv \frac{1}{(\alpha_1+2\alpha_2)}$  \cite{Rezende:2009sv}. This parameter plays a central role in the spectrum of quantum geometric operators, and controls, 
in the presence of Fermions, the strength of an emergent four-fermion interaction \cite{Perez:2005pm, Freidel:2005sn, Mercuri:2006um}.}. 

The previous facts motivate the idea of the pertinence of such variables for the implementation of non-perturbative quantization and thus can be viewed as natural rational behind the approach of loop quantum gravity \cite{Rovelli:2004tv} (although the history of the subject cannot be reduced literally to such perspective but rather to the discovery of Ashtekar's new variables \cite{Ashtekar:1986yd}). However, not surprisingly unlike the simple 3d case (which has no local degrees of freedom) the absence of ambiguities in the quantum theory remains an open question. 
Indeed, at early stages of the development of loop quantum gravity it was found that---thanks to the peculiar Hilbert space of quantum gravity adapted to diffeomorphism invariance and the Ashtekar-Barbero connection variables---the quantum gravitational dynamical equations (embodied by the Hamiltonian constraint that encodes both the gravity and matter interactions)
where free of UV divergences \cite{Thiemann:1996aw}. Nevertheless, the quantization of the Hamiltonian constraint suffers of ambiguities of an infinite dimensional nature suggesting that the renormalizability issue is still present \cite{Perez:2005fn}. 

There is however an unresolved consistency check concerning the quantization of the constraints in loop quantum gravity. This is the issue of anomalies. 
More precisely, the quantum dynamical equations are represented by a set of quantum operators that must satisfy a commutation algebra inherited from the classical algebra of generators of the surface deformation algebra. Checking the absence of  anomalies has shown to be a remarkably hard question suggesting that such consistency check could reduce the ambiguities in the definition of the quantum constraints \cite{Varadarajan:2018tei, Tomlin:2012ejk, Ashtekar:2020xll}. However, one should recognise that such hope is not clear from the general perspective of our initial discussion as the algebra of surface deformations is a feature of any diffeomorphism invariant formulation of gravity. More precisely, in the case of metric variables, the canonical analysis of the general action \eqref{actiong} would produce the same surface deformation algebra independently of the values of the undetermined couplings. 

The non-triviality of this question has motivated a recent interest in the application of  renomalization group methods to investigate (as in asymptotic safety scenarios) the possibility that the non-perturbative techniques of loop quantum gravity could help uncover a non trivial UV completion of the theory \cite{Bahr:2009qc,  Dittrich:2011zh,  Bahr:2014qza, Bahr:2011aa, Steinhaus:2020lgb, Thiemann:2020cuq}.  

The perspective that we emphasize here is not new \cite{Perez:2005fn} but its full implications in quantum cosmology has been somewhat underestimated. 
The problem of quantization ambiguities in the cosmological models inspired by the full theory has been considered in various works \cite{Ashtekar:2007em}; however, under certain restrictive assumptions that reduce the discussion to finite dimensional sectors of the space of ambiguities \cite{Vandersloot:2005kh, BenAchour:2016ajk, Dapor:2017rwv, Li:2018opr, Kowalczyk:2021bwr},  it has been argued not to represent a menace to the predictatbility of the framework. 
 Our present analysis shows that, as long as the question of ambiguities remains open in the full theory, polymerized symmetry reduced models cannot produce accurate quantitative predictions but only qualitative insights. We will show here that the ambiguities inherited from the full theory have an important dynamical effect in these models that, naturally,  compromises their predictive power. Nevertheless, due to their simplicity, our analysis does not reduce in any way the great value of these models for illustrating qualitative features of quantum gravity. Some of these features are, in these simple models at least, generic (i.e. independent of the ambiguity issue) suggesting that they might represent robust features possibly realized in nature.

\section{Unimodular quantum cosmology as a simple testing ground}\label{uuu}

In order to show the influence of the regularization ambiguities on physical quantities one first needs to be able to perform explicit calculations within the framework where these ambiguities appear. Even when quantum cosmology models correspond to classical systems with finitely many degrees of freedom, the non-standard representation theory used in the construction of the quantum theory makes these models sufficiently complicated (in some of its versions) to prevent explicit calculations. For example, different choices of time variables (realized by different choices of lapse functions) produce different quantum constraints which can present supplementary challenges when it comes to analysing the quantum dynamics.  

This is in part the reason why the technique of the so-called {\em effective dynamics} has been developed \cite{Taveras:2008ke} where the quantum evolution is approximated by modified classical equations of motion. 
These effective equations are affected by the ambiguities in the definition of the quantum dynamics. In fact these modifications are supposed to encode the quantum corrections to general relativity coming from quantum gravity. In this sense the quantization ambiguities on which we focus here are expected to affect these quantum corrections.  However, showing explicitly the form of the effective equations can be challenging or (unnecessarily) more involved when different time variables are chosen. In order to simplify our presentation we analyse cosmology in the unimodular version of general relativity. Unimodular gravity is simply equivalent to standard general relativity if the matter coupling is diffeomorphism invariant \cite{Ellis:2010uc}. When applied to cosmology, it has the advantage of resolving the problem of time as the lapse function is fixed by the unimodular constraint.  
It is customary in the literature of quantum cosmology to modify the scalar constraint by assuming different choices of the lapse function (a choice that it is often referred to as a `gauge choice'). Even when unimodular gravity is not a gauge fixing of standard general relativity, from the previous perspective unimodular cosmology could  be characterized  at the technical level  by a choice of lapse. We will see that such choice makes the Hamiltonian evolution particularly simple in the gravity sector, and thus allows for the most transparent and simple derivation and resolution of  the quantum as well as the effective dynamical equations.  It should be clear that the main point of this work will not change if one would use a different notion of lapse. 
The choice we make has a very natural geometric interpretation that we describe in the following paragraph. 

There is no preferred notion of time in general relativity. This implies that the dynamics is dictated by constraint equations and leads to the so-called problem of time in quantum gravity: instead of Schroedinger like evolution equations one has a timeless dynamics defined by the quantum constraints. This very complicated technical and conceptual problem can be circumvented in quantum cosmology  by the use of tools that have become customary in the area. The commonly accepted prescription is the use of some (partial) observable as clock that allows for the deparametrization of the dynamics that leads to a Schroedinger like evolution equation and the definition of the so-called physical Hilbert space of quantum cosmology. Even when such procedure is not unique and thus might lead to unitarily inequivalent theories, this additional source of potential ambiguity will not concern us here. The reason why the problem of time does not seem so serious in quantum cosmology is the fact that for the study of dynamical questions (sufficiently far form the Planckian regime) an effective classical description is available. Such classical description allows for dealing with the problem of time in just the usual way that is familiar to us in cosmology: via gauge fixing, i.e., particular choices with some clear interpretation ranging from co-moving time, harmonic time, conformal time, or (our choice here) unimodular time. 

Unimodular time is the time variable that naturally emerges from the description of cosmology in unimodular gravity \cite{Smolin:2010iq}. Instead of a gauge fixing, unimodular gravity can be thought of as a genuine modified theory of gravity that is (apart from a subtlety in the cosmological constant sector) is completely equivalent to general relativity.  We will use unimodular quantum cosmology \cite{Chiou:2010ne, Sartini:2020ycs} in what follows just because in this formulation the gravitational part of the Hamiltonian takes a particularly simple form; indeed, the geometry degrees of freedom can be mapped uniquely to those of a non-relativistic free particle. This feature allows for a very intuitive interpretation of both the effective classical evolution equations as well as the quantum gravity equations. In most situations of interest the problem of quantum or classical evolution of geometry coupled with simple forms of matter can be seen as a regular scattering problem of a non-relativistic particle in an external potential. This makes the setting of the dynamical system particularly appealing for its simplicity; however, it should be clear from our treatment that the implications drawn are of general validity and should apply (qualitatively speaking) to any of the customary parametrizations of loop quantum cosmology.

When specializing to (spatially flat)  homogeneous and isotropic cosmologies with metric
\be\label{frw}
ds^2=-N^2 dt^2+a(t)^2 d\vec x^2
\ee
the Einstein-Hilbert action supplemented with the unimodular constraint becomes
\be
S=-\kappa^{-1} \int \left(\sqrt{|g|} R +\lambda (\sqrt{|g|}-1) \right) dx^4
\ee
where $\kappa=16\pi G$, and  $\lambda$ is a Lagrange multiplier imposing the unimodular constraint $\sqrt{|g|}-1=0$, and we have put an overall minus sign in front of the action for later convenience. 
Specializing to the FLRW metric \ref{frw} one gets 
\be\label{uno}
S=\kappa^{-1}  V_0 \int\left( 6 \frac{a \dot a^2 }{N}-\lambda (N |a|^3-1) \right)dt,
\ee
where total derivative terms have been eliminated, and the $3$-volume $V_0$ of a fiducial cell has been introduced. 
Resolving the unimodular constraint fixes $N=|a|^{-3}$ and defines a preferred notion of time; from now on we denote this new time variable as $s$ and we call it unimodular time. The action becomes
\be
S=\kappa^{-1}  V_0 \int {6 a^4 \dot a^2 }ds,
\ee
where $s$ denotes from now on unimodular time. For further reference it is important to relate unimodular time with the standard comoving time $\tau$, namely
\be\label{tititime}
ds=- |a|^3 d\tau.
\ee
At this point we will change variables to more convenient ones that make the action look like that of a non-relativistic free particle.
The new configuration variable will be chosen to be given by the $3$-volume density
\be
x=a^3, 
\ee
from which it follows that $\dot x=3 a^2 \dot a$ and the action then is
\ba
\boxed{S=
 \int  \frac1{2} m\dot x^2 ds} ,
\ea
with 
\be
m\equiv \frac{4 V_0}{3  \kappa },
\ee
and where the dot denotes derivative with respect to the unimodular time $s$.
Note that the minus sign in from of \eqref{uno} was chosen so that the kinetic term of the particle analog has the usual sign. Also notice that if we use comoving time $d\tau= -ds/|a|^3$  we have 
\be
p=m \left(\frac{dx}{ds}\right) = -3 m  \frac{1}{|a|} \frac{da}{d\tau}=-3m \ {{\rm sign}(a)} H.
\ee
We see that the momentum variable in our parametrization is just proportional to the Hubble rate $H$ in usual comoving variables!
Let us introduce a scalar field as a matter model. Then the matter action (with the same overall minus sign convention that we are adopting) is
\ba\label{2dis}
S_M&=&\frac{1}{2}\int \sqrt{|g|} \left(\nabla_a\phi\nabla^a\phi+U(\phi)\right)\n \\ &=&-\frac{1}{2} V_0 \int N a^3 \left( \frac{1}{N^2} \left(\frac{d\phi}{ds}\right)^2-U(\phi)\right)ds= -\frac{1}{2} V_0 \int \left( a^6\left(\frac{d\phi}{ds}\right)^2-U(\phi)\right) ds.
\ea
Therefore, the action including our simple matter model is
\be\label{posta}
{S(x,\phi)=
 \int  \left(\frac1{2} m\left(\frac{dx}{ds}\right)^2-\frac{1}{2} V_0 x^2 \left(\frac{d\phi}{ds}\right)^2-V_0U(\phi)\right) ds} 
\ee
The previous action can be written in Hamiltonian form as
\be
{\boxed{S(x,\phi)=
 \int  p \frac{dx}{ds}+p_\phi \frac{d\phi}{ds} - \left(\frac{p^2}{2m}-\frac{p_\phi^2}{2 V_0 x^2}-V_0U(\phi)  \right) ds } },
\ee
where \be p_\phi=-V_0 x^2 \frac{d\phi}{ds}.\ee

\subsubsection{Changing variables to match the standard setup}

The previous $(x,p)$ variables in the gravity sector where chosen to emphasize the simple link between unimodular gravity in the FLRW context and the dynamics of a point particle. A simple rescaling of these variables leads to the standard parametrization of the phase space in loop quantum cosmology in the so-called $\bar\mu$-scheme. The variables customarily used are called $(b,v)$ and are defined as
\be
b\equiv  -\frac \gamma {3m} p=-\frac{\gamma}{3} \frac{d x}{ds} =-\gamma a^2 \frac{d a}{ds},
\ee
and
\be
v\equiv \frac{3 m}{\gamma} x =\frac{V_0 a^3}{4\pi G\gamma}.
\ee
Thus one has
 \be
\{b,v\}=1.
\ee
With these variables, the Hamiltonian is
\be\label{hammm}
H=\frac{V_0}{2\pi G \gamma^2}\left(\frac{3}{4} b^2-\frac{p_\phi^2}{16 \pi G v^2}-2\pi G \gamma^2\, U(\phi)\right),
\ee
which is proportional to the scalar constraint $C$ as written in reference  \cite{Ashtekar:2011ni} (equation 2.19) simply rescaled by the use of the unimodular time lapse, namely $H=V_0 C/(\pi G |v|)$. The advantage of using unimodular variables resides in the remarkable fact that the gravity part of the Hamiltonian depends only on the variable $b$ (like a free particle in classical mechanics). This simple fact simplifies several technical as well as conceptual discussions of the classical and quantum features of the model.

\section{Regularization ambiguities of the Hamiltonian}

There are two aspects of the Hamiltonian that call for a modification of its classical expression in order to
promote it to a well defined self-adjoint operator in the special Hilbert space  of loop quantum cosmology. 
One of them is that  only quasi-periodic functions  
 of $b$ but not $b$ itself can be quantized. The second is that inverse volume contributions to the Hamiltonian (entering through the matter coupling)
 are also modified by means of the use of classical expressions that eliminate unboundedness of these at small volumes. Both modifications are ambiguous by nature and lead to dynamical effects that we analyze in what follows. Interest in this issue from the observational perspective has resurfaced recently in \cite{Renevey:2021tmh}, here we concentrate on further theoretical implications. There are other approaches for the definition of the quantum dynamics for cosmology where one starts from a more fundamental perspective at the quantum level and infers from it the symmetry reduced model \cite{Gielen:2016dss, Oriti:2016qtz, Alesci:2013xd}, we note that similar ambiguities are present in these perspectives as well. For simplicity we concentrate on the loop quantum cosmology formulation where the problem is embodied in the notion of regularization.

\subsection{Holonomy corrections}\label{holycow}

Due to the peculiar choice of representation in the quantization of the model (inspired by the structure of Loop Quantum Gravity), there is no $b$ operator in the Hilbert space of loop quantum cosmology but only operators corresponding to finite $v$ translations \cite{Ashtekar:2006wn, Ashtekar:2011ni}; from here on referred to as shift operators \be\label{shifty} {\exp(i \lambda b)}\triangleright \Psi(v )=\Psi(v-\lambda),\ee 
where $\lambda$ is some arbitrary length scale. As the classical Hamiltonian explicitly depends on $b$, it needs regularization in order to be promoted to a self-adjoint operator in the 
Hilbert space of loop quantum cosmology. Consequently we replace \eqref{hammm} by 
\be\label{Hamon}
H=\frac{V_0}{2\pi G\gamma^2}\left(\frac{3f(\lambda b)^2}{4\lambda^2}-\frac{p_\phi^2}{16 \pi G v^2}-2\pi G \gamma^2 \, U(\phi)\right),
\ee
where
\be\label{only-condition}
f(\lambda b)=\sum_{n\in \Z} f_n e^{in\lambda b}.
\ee
To be explicit about the regularization choice we write the substitution rule as
\be\label{regularis}
\boxed{b^2\to \frac{f(\lambda b)^2}{\lambda^2}.}
\ee
This operation is called {\em polymerization} in the loop quantum cosmology literature.
The only conditions that consistency with the classical dynamics imposes are: On the one hand,  that $f(x)=\bar f(x)$ which translates into the condition
\be
f_n=\bar f_{-n}.
\ee 
On the other hand one needs that \be \label{naive}\braket{f(\lambda b)}=\lambda b_0+\sO[(\lambda b_0)^2]\ee for $\lambda b_0\ll 1$ when expectation values are computed in suitable semiclassical states peaked at the classical momentum $b_0$. 
This second condition is necessary to recover the semiclassical dynamics of standard cosmology at low Hubble rates or the low density regime. 
This leaves an infinite dimensional freedom in the choice of the regularized Hamiltonian to be promoted to an operator in the Hilbert space of our system. The standard choice in the loop quantum cosmology literature is $f_n=i\delta_{n}^{1}/2$, namely
\be\label{tradition}
b^2\to \frac{\sin^2(\lambda b)}{\lambda^2}.
\ee
A possible justification for this choice is that of simplicity. The link between such choice in relation to the lowest non-vanishing eigenstate of the area operator in loop quantum gravity, and the special status given to  the fundamental representation of the gauge group, is sometimes evoked as a further reason to pick \eqref{tradition} (See \cite{Ashtekar:2011ni}). This argument connects the regularization of a certain quantum operator in loop quantum cosmology to the features of a particular state (the state with minimal area-eigenvalue) in loop quantum gravity. Even when accepting such possibility it is unclear how the lowest area eigenstate should play such a central role. Indeed, in quantum theory the principle of superposition rather suggests that states would typically be made of arbitrary superpositions of different area eigenvalues.  Consideration of such aspects in full generality brings us back to the infinite dimensional landscape of {\em polymerizations} in \eqref{regularis}.

When expressed in terms of the $v$-basis the evolution equation (related to the Hamiltonian constraint) contains a finite difference term, which, with the so-called {\em traditional choice} \eqref{tradition} 
becomes a discrete version of a second derivative in $v$. For an arbitrary choice \eqref{regularis} the finite difference term can be put in correspondence with a linear combination of
the discretization of higher derivative terms in $v$. If one were looking for eigenstates of the Hamiltonian (\ref{Hamon}) one would be confronted with a growing multiplicity of formal solutions of the eigenvalue equation as the order of the corresponding difference equation grows when considering general functions $f(b)$ with arbitrarily high Fourier components. This question can be studied in the simpler context of the pure gravity case which in the analogy with the point particle system corresponds to the asymptotic large universe regime where standard matter contributions can be neglected in a logic analogous to that of scattering theory. In such simpler setting, most of these extra solutions related to the higher order character of the difference equation for arbitrary $f(b)$  can be shown not to be normalizable, and hence not to be part of the spectrum.

In the Wheeler-DeWitt standard representation of quantum cosmology, eigenstates of the Hamiltonian are doubly degenerate in correspondence with the two possible equal `energy' classical solutions corresponding to an expanding and/or contracting universe for a given cosmological constant: the two are related to the discrete symmetry $\dot x\to -\dot x$ involving the initial conditions of the theory written in the variables \eqref{posta}.  Therefore, any additional degeneracy of energy eigenvalues would have no classical correspondence and its associated conserved quantity would reveal the existence of new (microscopic) degrees of freedom. We will see that this is the case in a subtle way but we postpone this discussion until Section \ref{qqq}.

The formal similarities with higher curvature corrections of the Einstein-Hilbert action due to quantum effects is manifest even when a rigorous statement in this sense is made difficult by the breaking of explicit covariance by the Hamiltonian formulation (in the first place), and the further (possibly explicit) breaking of covariance introduced by  the polymerization itself (see \cite{Bojowald:2015zha}). In perturbative quantum gravity, higher derivative terms arise from higher curvature corrections and this changes the number of degrees of freedom as seen from a classical perspective. 
As higher curvature (higher derivative) terms appear multiplied by increasing powers of $\ell_p^2$, all these corrections are taken to be negligible at energy scales well below the Planck scale. 
We believe that this analogy is interesting, thus  making polymer models a nice simplified arena where difficult questions related to renormalization and the definition of the continuum limit can be explored in the highly simplified context of a model that (at least classically) starts with a finite number of degrees of freedom.
   
Finally, it is possible to exhibit the direct relation between the function $f(\lambda b)$ and the cosmological constant as follows. Standard considerations in unimodular gravity imply that \cite{Amadei:2019wjp, Amadei:2019ssp}
\be\label{Lambda}
\Lambda=8\pi G  \frac{E}{V_0},
\ee
where $E$ is the eigenvalue of $\eqref{hammm}$ or $\eqref{Hamon}$. 
The discussion is simplified if we assume that we are in the massless scalar field case $U(\phi)=0$. If a non trivial self interaction is present then a more careful analysis is needed. We restrict to initial conditions given at $v=\pm \infty$ (large universes) where the contribution of the scalar field to the Hamiltonian \eqref{Hamon} vanishes. In this limit the system is the analog of a non relativistic free particle where eigenvalues of the energy can be labelled by eigenvalues of momenta\footnote{More precisely, in our context these correspond to the eigenvalues of the shift operators \eqref{eige}, yet the key point is that they are still labelled by a value of $b$.}. Therefore, one obtains from \eqref{Hamon} the relation
\be
\Lambda=\frac{3 f^2(b_\infty)}{ \gamma^2 \lambda^2},
\ee
where $b_\infty$ is the asymmptotic value of $b$ for $v=\pm \infty$.

%
%
%
\subsection{Inverse volume corrections}\label{invv}

\begin{figure}
\centering
\includegraphics[width=0.8\linewidth]{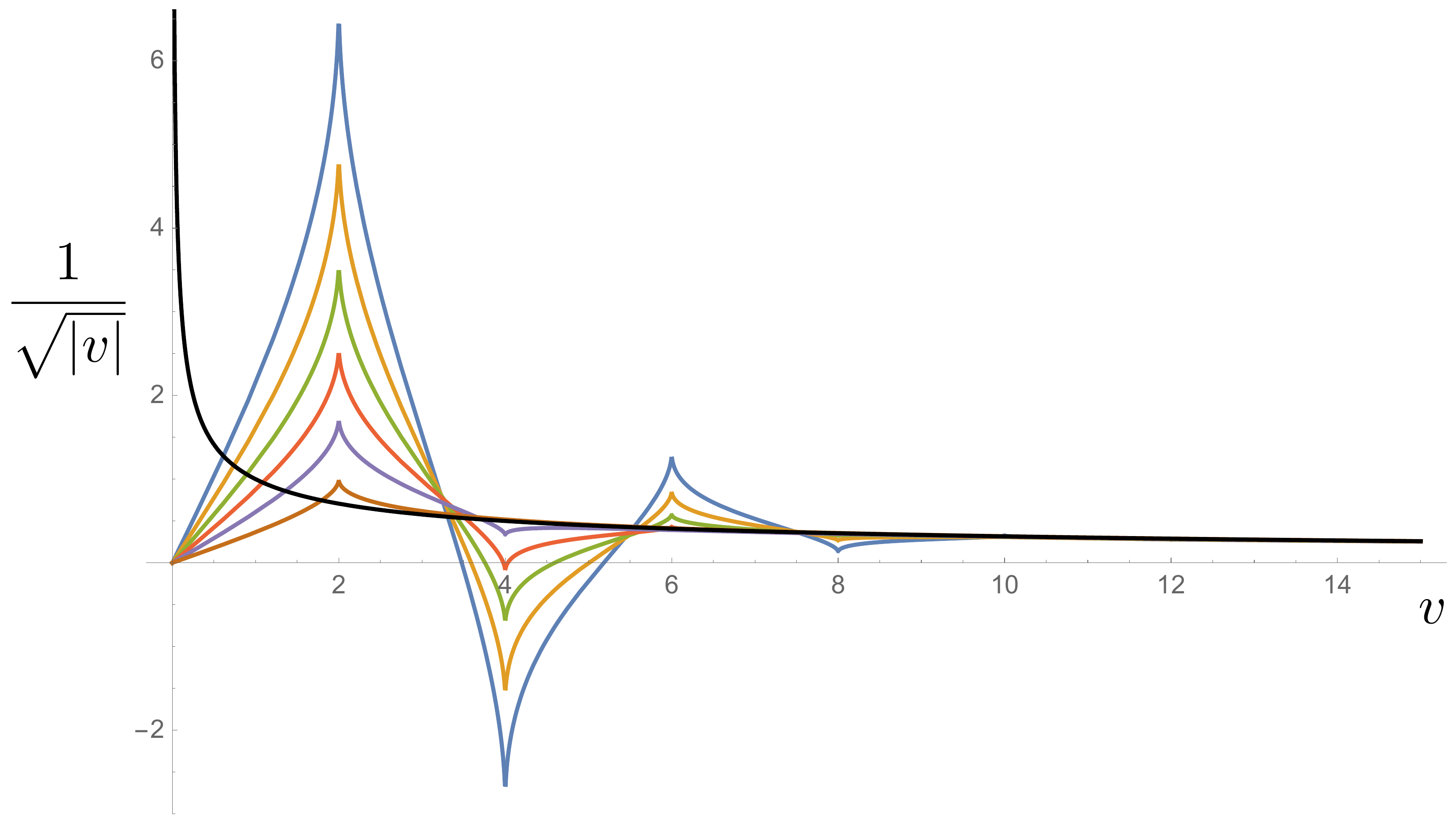} 
\caption{Inverse volume corrections of the classical function $1/\sqrt{|v|}$ (shown in black). The blue curve represents the best approximation for large volumes which coincides with the classical expression up to order $(\lambda/v)^{10}$. However, as the approximations get better for large volume they get worse at Planckian volumes as the plotted sequence illustrates which deviations becoming the largest at low integer values times $\lambda$. } \label{inverse-volume}   
\end{figure}

Inverse volume terms in the Hamiltonian introduce potential singularities in the quantum theory. Such potential divergencies are present as well in the full theory of loop quantum gravity 
and need regularization when constructing a well defined quantum scalar constraint operator. Thiemann introduced \cite{Thiemann:1997rt}  a natural regularization of such potential UV divergences by realizing that inverse volume terms can be obtained from the Poisson algebra between well defined geometric operators and the holonomy of the connection. In the case of cosmology the idea can be illustrated, for example, by the following simple classical identity\footnote{This is a particular case of a more general identity leading to additional ambiguities \cite{Singh:2013ava}. For simplicity we concentrate on the one given in \cite{Ashtekar:2011ni}.}  
\be\label{prima}
\frac{1}{\sqrt{|v|}}= \frac{2i}{ \lambda}{\rm sgn}(v) \exp{(i\lambda b)}\ \{\exp{(-i\lambda b)},\sqrt{|v|}\}
\ee
which suggests a natural regularization of quantities depending on the inverse volume using `holonomies' and commutators in the quantum theory. Using a symmetrized factor ordering, for instance
\be
\widehat{\frac{1}{\sqrt{|v|}}}\to  \frac{1}{ \hbar \lambda}{\rm sgn}(v)\left( \exp{(i\lambda b)}\ [\exp{(-i\lambda b)},\widehat{\sqrt{|v|}}]+ [\exp{(-i\lambda b)},\widehat{\sqrt{|v|}}]\  \exp{(i\lambda b)}\right)
\ee
This choice regularizes the singular behaviour of the inverse volume at $v=0$---where the previous expression vanishes by construction---and produces a well defined operator in the Hilbert space of loop quantum cosmology. However, the choice is by no means unique. In fact (in addition to factor ordering and other sources of ambiguities, such as the choice of the power of $v$ inside the Poisson bracket in \eqref{prima}) on has an infinite dimensional space of regularizations that is similar in spirit to the one identified for the regularization of curvature in \eqref{Hamon} given by
\be
\widehat{\frac{1}{\sqrt{|v|}}}\to  \frac{{\rm sgn}(v)}{ 2\hbar  \sum_{m\in \Z} c_m} \sum_{n\in \Z} \frac{c_n}{ \lambda n}\left( \exp{(i\lambda n b)}\ [\exp{(-i\lambda n b)},\widehat{\sqrt{|v|}}]+ [\exp{(-i\lambda  n b)},\widehat{\sqrt{|v|}}]\  \exp{(i\lambda n b)}\right),
\ee
for arbitrary coefficients $c_n$. This implies that in addition to the infinite dimensional family of curvature regularizations one has an (at least) equally large family of inverse volume regularizations which would generically enter in the construction of the matter coupling when defining the quantum Hamiltonian. One can show that the action of the previous operator is simply given by  (see \cite{Ashtekar:2011ni})
\ba\label{reguinv}
\widehat{\frac{1}{\sqrt{|v|}}}\Psi(v) & =& \frac{\Psi(v) }{ \hbar \sum_{m\in \Z} c_m} \sum_{n\in \Z} \frac{c_n}{  \lambda n}\left(\sqrt{\left|v+\lambda n\right|}-\sqrt{\left|v-\lambda n\right|} \right)\n \\
&\equiv& \frac{\Psi(v) }{ \sum_{m\in \Z} c_m} \sum_{n\in \Z} {c_n} \left[\frac{1}{\sqrt{|v|}}\right]_n,
\ea
where we have introduced the definition 
\be\label{decho}
\left[\frac{1}{\sqrt{|v|}}\right]_n\equiv\frac{1}{ \hbar \lambda n}\left(\sqrt{\left|v+\lambda n\right|}-\sqrt{\left|v-\lambda n\right|} \right).
\ee
One has that
\be
\sqrt{|v|} \left[\frac{1}{\sqrt{|v|}}\right]_n=1+\frac{1}{16}\frac{n^2\lambda^2}{v^2}+\frac{7}{128}\frac{n^4\lambda^4}{v^4}+\sO\left(\frac{n^6\lambda^6}{v^6}\right), 
\ee
which shows that for sufficiently large volume one recovers the classical expected limit. Notice that the regularization \eqref{reguinv} vanishes at $v=0$. 
One can use the previous series expansion and chose the coefficients $c_n$ in order to improve the convergence to the classical value. For example with the following choice
\be
\text{c1}\to 9.42267, \ \text{c2}\to -13.1273, \ \text{c3}\to
   6.31659, \ \text{c4}\to -1.93791,\ \text{c5}\to 0.355751,\ \text{c6}\to
   -0.0297957
\ee
one gets the regularization to coincide with $1/\sqrt{|v|}$ up to order $\sO\left({n^{10}\lambda^{10}}/{v^{10}}\right)$ (plotted in blue in Figure \ref{inverse-volume})! One can continue improving convergence by killing higher order deviations from $1/\sqrt{v}$, it simply boils down to solving a linear system of equations with increasing dimension. One might think that such process would produce a sequence of $c_n$ converging pointwise to the Wilson-Ewing regularization \cite{Wilson-Ewing:2012dcf, Singh:2013ava} of the inverse volume which is given by $1/\sqrt{|v|}$ for all $v\not=0$ while it vanishes at $v=0$. By plotting a few members of the above approximating sequence we see that this will not be the case. In fact, the previous sequence, while it gets better and better in approximating  $1/\sqrt{v}$ for large $v$, it differs more and more from the function $1/\sqrt{v}$ at around $v=\pm \lambda$ (see Figure \ref{inverse-volume})


 There are two important features that we would like to emphasize here. The first one is that the regularization of the inverse volume operator would lead to a function of $v$ that is not everywhere differentiable due to the presence of the absolute value in the formulas.  This is of course not a problem from the perspective of the quantum theory where (for the quantum dynamics) only the evaluation of the regularization on a discrete lattice plays a role. However, at the non differentiable points the effective equations can simply not be trusted. The second important feature is that the regularization is a continuous function and thus bounded.  We insist on the point that (even when some choices might seem natural given some subjective criteria) there is no well defined rule that would actually eliminate the vast set of possibilities here either.  We will discuss some further consequences of inverse volume corrections later, in particlar we will see in  Section \ref{nec} their role in the 
violation of the {\em null energy condition} near the big bang.


\section{The quantum theory: evolution across the singularity}\label{qqq}

Let us discuss the main features of the quantum dynamics, before discussing the validity of the effective dynamical approach that we will use later for further interpretation---where the quantum dynamics is approximated by looking at the evolution of semiclassical states. One great advantage of the unimodular gravity formulation is that (at least in the FLRW context) the theory has a well identified time evolution (in unimodular time \eqref{tititime}) and the Kinematical Hilbert space of loop quantum cosmology is the physical Hilbert space. In other words the problem of time is trivialized and the physical interpretation of the quantum theory becomes closer to that of standard quantum mechanics \footnote{Closer but not quite exactly the same. Here we are making reference to the unconfortable questions related to the meaning of a quantum theory of the universe as a whole. These questions are indeed very important and remain open to a large extent. In order to concentrate on the main point of this paper we have to ignore them altogether.}. However, an important difference with standard quantum mechanics is the use of an unconventional  representation of the basic phase space variables which brings into the system a central property of the full theory of loop quantum gravity: fundamental discreteness. Concretely, instead of the standard Schroedinger representation on a gravitational Hilbert space of square integrable wave functions where $v$ acts by multiplication and $b=-i\partial_v$, one introduces a Hilbert space where only the exponentiated version of $b$, the shift (or holonomy) operators $\exp{i\lambda b}$ (for arbitrary $\lambda\in \R$) are well defined. This is called sometimes the \emph{polymer representation} and the procedure of using this exotic representation (well motivated from the full theory) is called {\em polymerization}.

Consequently,  there is no operator corresponding to $b$ in the loop quantum cosmology polymer representation but only the operators corresponding to finite $v$ translations \cite{Ashtekar:2006wn}; from here on referred to as shift operators defined as \be\label{shifty} {\exp(i \lambda b)} \triangleright\Psi(v )=\Psi(v-\lambda).\ee 

There are states diagonalizing the shift operators, denoted $\ket{b_0; \Gamma^{\epsilon}_\lambda}$, which are labelled by a real value $b_0$ and where  $\Gamma^{\epsilon}_{\lambda}$  is a 1d lattice of points, a graph, in the real line of the form $v=n \lambda+\epsilon$ with $\epsilon\in [0,\lambda)$ and $n\in \N$. The corresponding  wave function of these eigenstates is given by $\Psi_{b_0}(v)\equiv \braket{v|b_0; \Gamma^{\epsilon}_\lambda}=\exp{(-i{b_0 v})} \delta_{\Gamma^{\epsilon}_\lambda}$ where the symbol $\delta_{\Gamma^{\epsilon}_\lambda}$ evaluates to one when  $v\in \Gamma_\lambda^\epsilon$ and vanishes otherwise. Assuming that $k=m\lambda$, it follows from (\ref{shifty}) that
\be\label{eige}  {\exp(i  k b)} \triangleright \ket{b_0; \Gamma^{\epsilon}_{\lambda}}=
\exp{(i k b_0)} \ket{b_0; \Gamma^{\epsilon}_{\lambda}}.\ee
 The states $\ket{b; \Gamma^{\epsilon}_\lambda}$ are eigenstates of those shift operators which preserve the lattice $\Gamma^{\epsilon}_\lambda$. The fact that these states are supported on discrete lattices (polymer-like excitations)  is what motivated the name of the representation. Notice that the eigenvalues are independent of the parameter $\epsilon$,  i.e.,  they are infinitely degenerate and span a non separable subspace of the quantum cosmology Hilbert space $\sH_{lqc}$. 

As the operator $b$ does not exist in the Hilbert space one has to construct approximations in terms of combinations of shift operators 
which behave like $b$ in a suitable sense. A procedure that is (as discussed in Section \ref{holycow}) intrinsically ambiguous.
We would like to understand the influence of deviating from the standard regularization \eqref{tradition} to the quantum dynamics. 
In order to do this we will concentrate on the pure gravity case first. Indeed, the ambiguity \eqref{regularis}  only affects the gravitational part of the Hamiltonian and thus this simple case will completely characterize the dynamical influence of the choice of different regularization functions $f(\lambda b)$ in the quantum dynamics in the large volume asymptotic regime where matter dilutes until becoming negligible. Thus we will deal with the special case where, before quantization,  the classical Hamiltonian is regularized as 
\begin{equation}\label{eq40}
    H=\frac{b^2}{2m}\ \ \ \to \ \ \ \frac{f(\lambda b)^2}{2m\lambda^2}.
\end{equation}
Note that this case is non trivial because it admits a non-zero cosmological constant that is given by the value of the energy in the unimodular framework (recall \eqref{Lambda}).
In the quantum theory, we are interested in the eigenstates of the Hamiltonian (the analog of the time independent Schroedinger equation). Let us first analyze the spectrum of the Hamiltonian in the traditional polymerization, namely we would like to solve the equation
\be
\left(\frac{\sin(\lambda b)^2}{2m\lambda^2}-E\right)\ket{\Psi_E}=0
\ee
which in the $v$-basis becomes (due to \eqref{shifty}) the difference equation
\be
\Psi_E(v-2\lambda)+\Psi_E(v+2\lambda) + (8m\lambda^2 E-2) \Psi_E(v)=0,
\ee
where the order of the difference equation is directly related to the polymerization choice. This seems to raise a potential difficulty: if instead of the traditional choice we take an arbitrary $f(\lambda b)$ the order of the difference equation will grow arbitrarily. Would this not lead to an uncontrollable proliferation of spurious solutions?  We will see soon that this is not the case. For the moment we continue the analysis of the present scenario. As the Hamiltonian is a combination of shift operators \eqref{shifty} of the kind for which one knows the eigenstates, one can simply express the energy eigenstates in terms of $\ket{b_0; \Gamma^{\epsilon}_{2\lambda}}$ (the eigenstates of the shift operators) and calculate the relationship between $b_0$ and the energy eigenvalues.  We could call this the  {\em  polymerized dispersion relations}. For the standard choice energy eigenstates and dispersion relations are 
\be\label{dispersion}
\ket{\Psi_{E(b_0)}}= \ket{b_0; \Gamma^{\epsilon}_{2\lambda}},\ \ \ \ \ \ E(b_0)=\frac{\sin(\lambda b_0)^2}{2m\lambda^2}.
\ee

\subsection{The $\epsilon$-sectors}

The previous energy eigenstates (eigenstates of the cosmological constant) are infinitely degenerate due to the $\epsilon$ degeneracy of the shift operators \eqref{shifty}. This over abundance of solutions of the Schroedinger equation is controlled, in standard accounts, by fixing a volume lattice once and for all and choising one $\epsilon$-sector. 
This choice is dynamically consistent because the Hamiltonian preserves the given lattice; however, in the presence of matter the choice represents an additional dynamical ambiguity as the dynamical features will depend on $\epsilon$. This is particularly clear when we look at the inverse volume corrections of Section \ref{invv}. Each different choice of $\epsilon$ gives a lattice that probes the volume regularization at different discrete points. As the inverse volume regularization enters the coupling of gravity with matter (see for instance equation \ref{Hamon}) the details of the dynamics will depend on this choice.  Because the Hamiltonian preserves $\epsilon$-sectors they are some times called {\em super selection} sectors. However, as there are other (Dirac) observables that do not preserve the lattice, these sectors are not superselected in any usual sense. 

More precisely, in the case of pure gravity observables commuting with the Hamiltonian and mapping between different values of $\epsilon$ (graph changing observables) are simply the shift operators introduced in \ref{shifty}. In the case of a non trivial matter coupling other Dirac observables exist, they are technically hard to characterize explicitly in their full generality because of the usual difficulty associated to the construction of such conserved quantities. However, notice that if the matter coupling is such that matter dilutes as $v\to \infty$ (as expected for regular matter degrees of freedom) then shift operators remain Dirac asymptotic observables where the universe becomes large and the Hamiltonian tends to the pure geometry Hamiltonian in the usual sense of scattering theory. 
The shift observables \eqref{shifty} define in this manner a {\em complete set of commuting observables} fully characterizing the positive energy (positive cosmologial constant\footnote{Negative $\Lambda$ solutions exist in the presence of matter and they correspond bound states (in the analog  non-relativistic particle system). These solutions do not reach the $|v|\to \infty$ asymptotic region and admit no scattering theory interpretation (as in usual cases).}) states.  These asymptotic observables are like those regularly employed in standard situations involving scattering theory. Their existence shows that $\epsilon$-sectors are not superselected.

There is hence no clear reason to restrict to a single lattice and superpositions of different lattices can be considered. Some of the implications of this possibility have beed investigated in \cite{Amadei:2019wjp, Amadei:2019ssp} where it is shown that these additional degrees of freedom, which are microscopic or Planckian, can be key in understanding the fate of information in situations where evolution across would-be-singularities is relevant like in cosmology and (most importantly) in the context of black hole formation and evaporation. In order to simplify the following discussion, we will restrict, from here on, our analysis to the case of states supported a single lattice.  

\subsection{Degeneracy of the energy (cosmological constant) eigenstates in the pure gravity case}

\begin{figure}[h]  
\centerline{\hspace{0.5cm} \(
 \begin{array}{c}
\includegraphics[width=7cm]{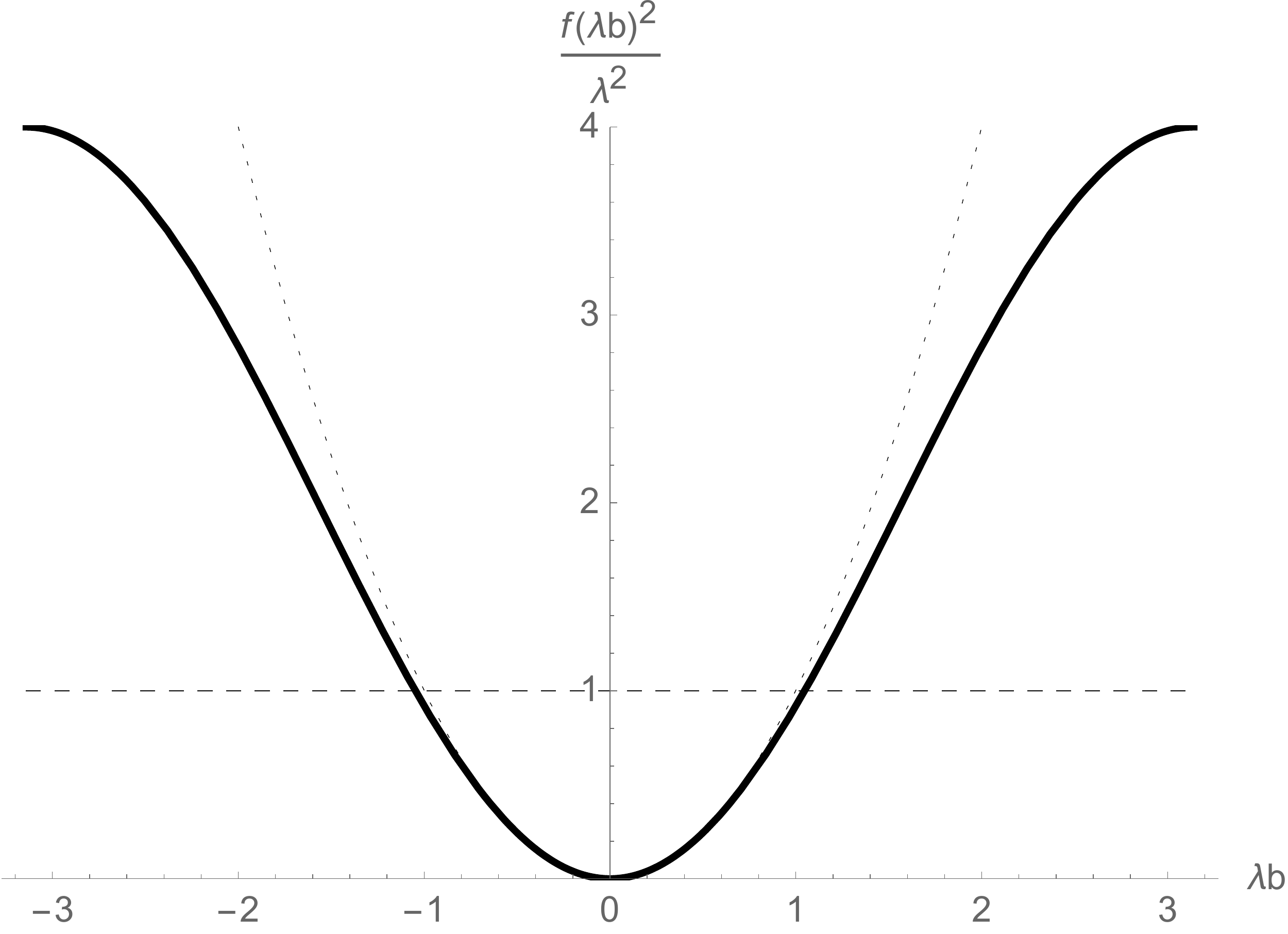} 
\end{array} \ \ \ \ \ \ \ \ \ \ \ \ \ \ 
 \begin{array}{c}
\includegraphics[width=9cm]{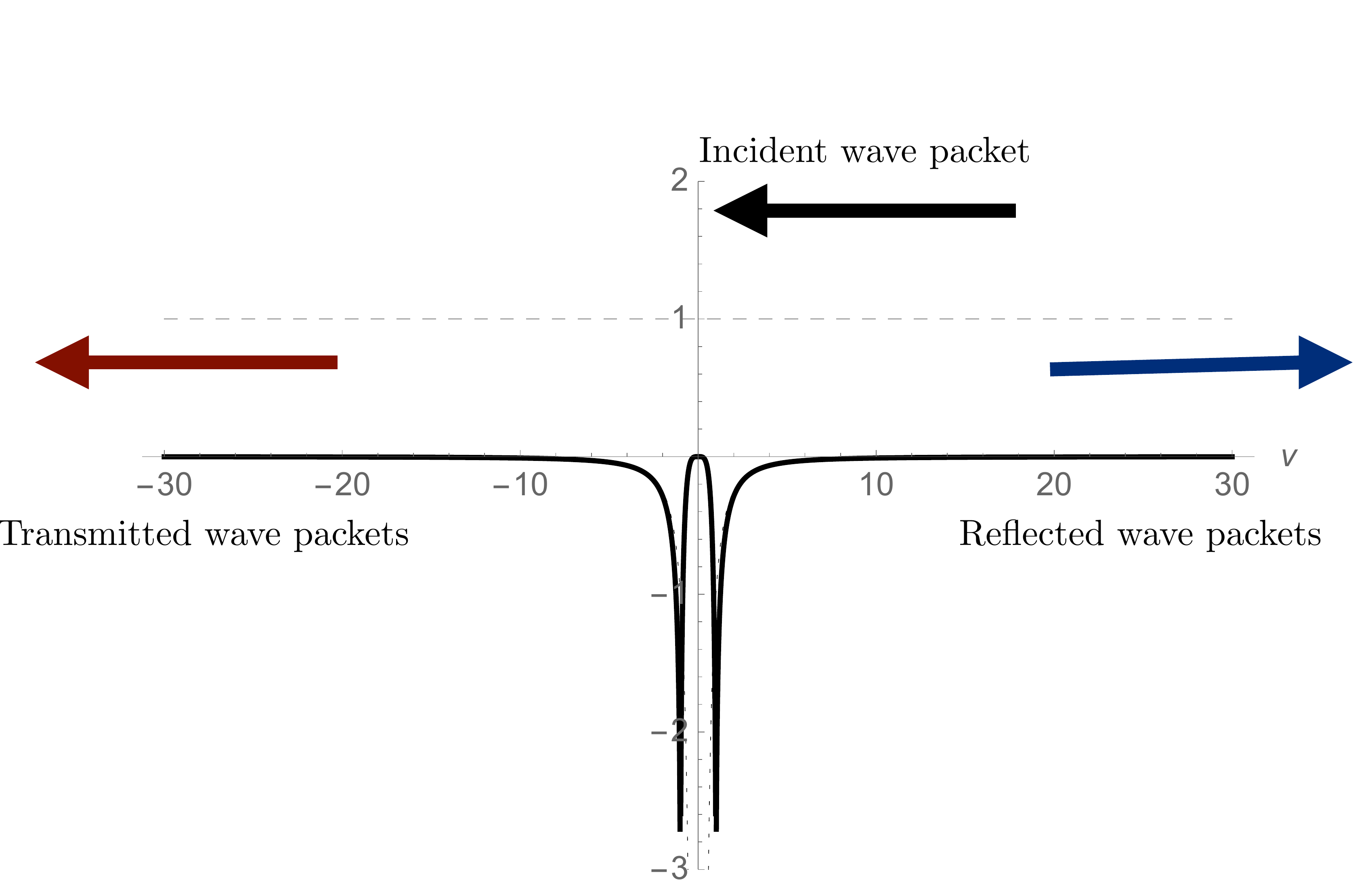} 
\end{array} 
\)} 
\caption{The polymerization $f(x)^2=-2(\cos(x)-1)$ matches the degeneracy of eigenstates of the Schroedinger representation. On the right panel we show the different scattering channels in a matter coupling with a massless scalar that produces (for a given $p_{\phi}$ eigenstate)
an effective potential regularized by inverse volume corrections (shown in black). 
 The universe bounces into a superposition of transmitted and reflected modes with the same asymptotic (large $v$) Hubble rates. If we factor by the symmetry $v\to-v$ then we only have a bounce and the superposition disappears.  The results of an analytically solvable model are shown in \eqref{QS}.} \label{diff1}  
\end{figure}

In the Schroedinger representation the dispersion relations would have been the familiar non relativistic particle relation $E(b_0)=b_0^2/(2m)$ which is doubly degenerate corresponding to the momentum eigenstates with $b=\pm b_0$. Translating this to unimodular cosmology, these two eigenstates would correspond to a state of a De Sitter with cosmological constant $\Lambda=8\pi G E(b_0)/V_0$ that is either contracting or expanding in the FLRW slicing. 
With the standard polymerization \eqref{tradition} we observe at first that a new degeneracy has appeared as there are four different shift-operator-eigenstates that produce the same energy, namely those labelled by the four roots of the equation on the right of \eqref{dispersion} depicted on the left of Figure \ref{diff1}. 
We will study the role of these additional solutions below once we have described this type of degeneracy for an arbitrary regularization.

\begin{figure}[h]  
\centerline{\hspace{0.5cm} \(
 \begin{array}{c}
\includegraphics[width=7cm]{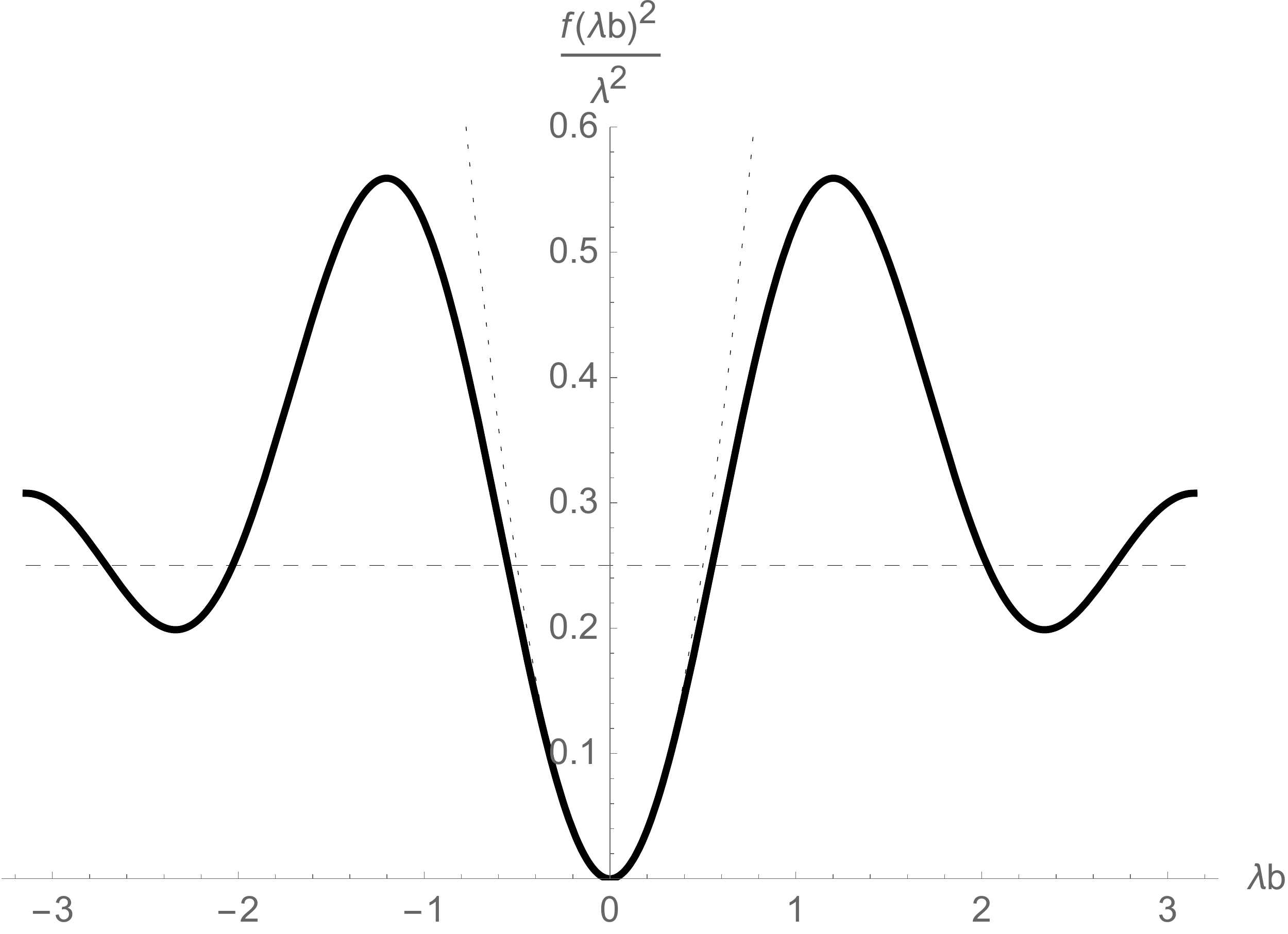} 
\end{array} \ \ \ \ \ \ \ \ \ \ \ \ \ \ 
 \begin{array}{c}
\includegraphics[width=9cm]{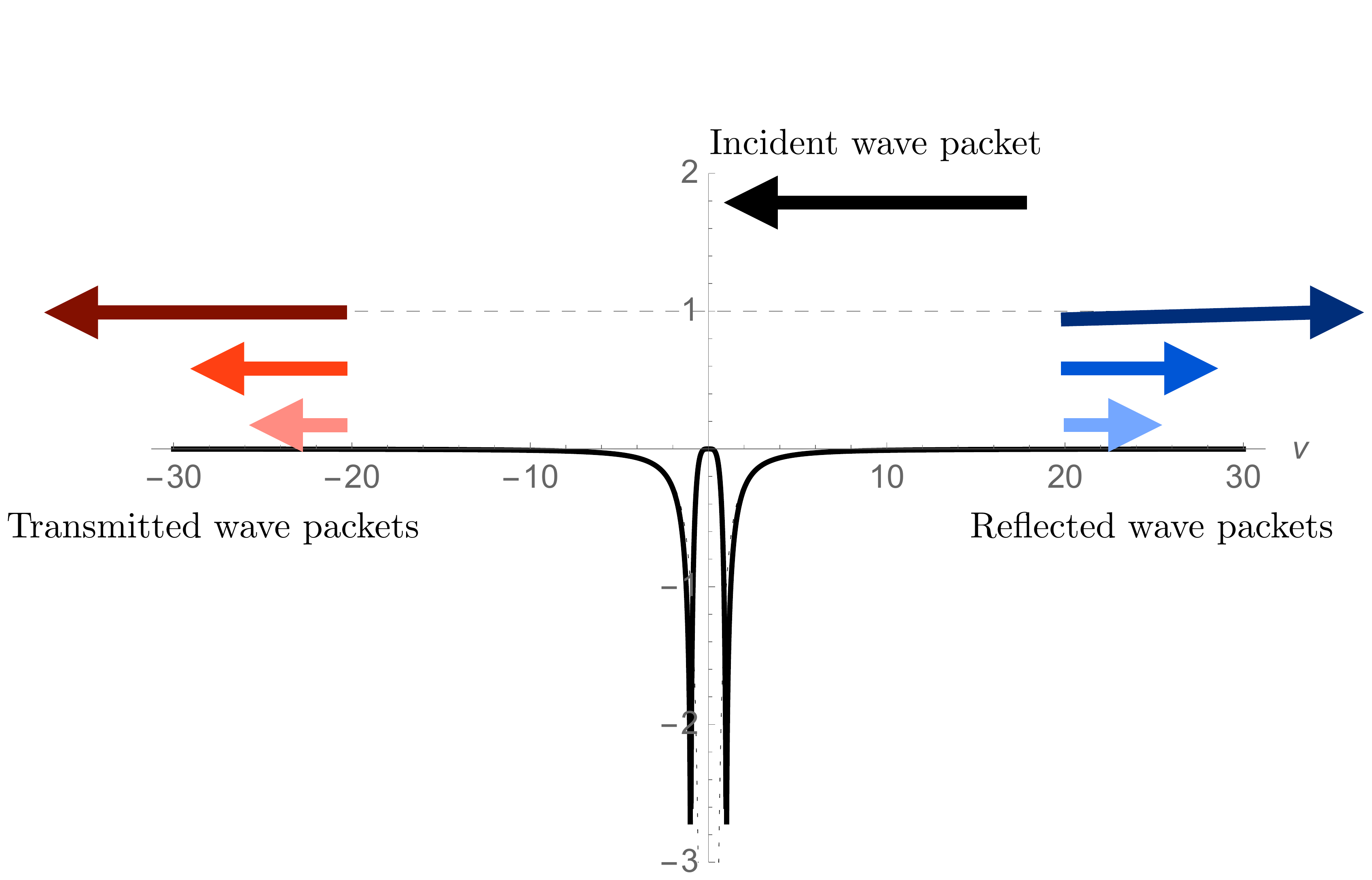} 
\end{array} 
\)} 
\caption{In a generic polymerization, here $f(x)^2=2/13(2 -(\cos[2 x] + \cos[3 x]))$. On the right panel we show the different scattering channels in a matter coupling with a massless scalar that produces (for a given $p_{\phi}$ eigenstate)
an effective potential regularized by inverse volume corrections (shown in black). 
New channels for the bounce appear and a given universe evolves through the singularity into a quantum supperposition of universes with the same cosmological constant (or expectation value of the cosmological constant for wave packets) but different Hubble rates.  The results of an analytically solvable model are shown in \eqref{QS}.} \label{diff3}  
\end{figure}

For an arbitrary polymerization \eqref{regularis} the situation is quite similar. Eigenstates are again given by 
\be\label{dispersion-bis}
\ket{\Psi_{E(b_0)}}= \ket{b_0; \Gamma^{\epsilon}_{2\lambda}}\ \ \ {\rm with}\ \ \ E(b_0)=\frac{f(\lambda b_0)^2}{2m\lambda^2}.
\ee
However, the degeneracy of the energy eigenvalues is now dependent on the choice of the function $f(\lambda b)$. An example with 6 different eigenstates is depicted on the right panel of Figure \ref{diff3}. One can distinguish in this example two different situations, one where the energy is $E_1$ and the 8 solutions correspond to eigenstates of the form \eqref{dispersion-bis}. In the other case, for the energy $E_2$, the number of solutions of the eigenvalue equation seen as a difference equation remains 8; however, only the four values of $b_0$ explicitly seen in the figure correspond to `plane-wave' eigenstates of the form \eqref{eige}. It is easy to show\footnote{See for instance section 2.3 of \cite{10.5555/230196}.} that the other 4 solutions of the difference equation are diverging in either the $v\to \pm \infty$ limit and thus are not part of the Hilbert space (this is the analog of non-normalizable solutions of for example the time independent Schroedinger equation for the Harmonic oscillator). But the most interesting thing concerning these additional solutions happens when matter couplings that break $v$-translational invariance of the Hamiltonian are included.

\begin{figure}[h]  
\centerline{\hspace{0.5cm} \(
 \begin{array}{c}
\includegraphics[width=7cm]{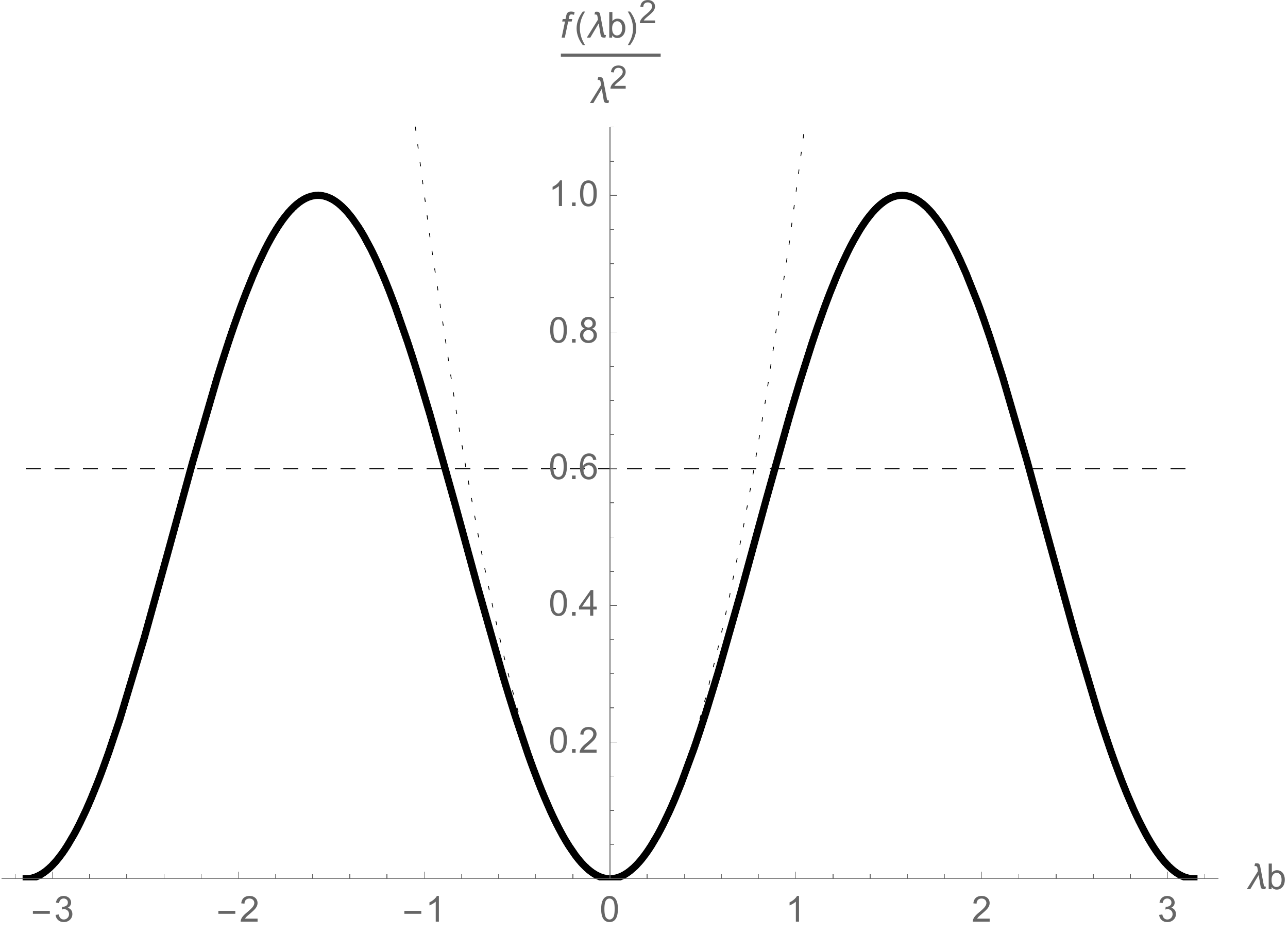} 
\end{array} \ \ \ \ \ \ \ \ \ \ \ \ \ \ 
 \begin{array}{c}
\includegraphics[width=9cm]{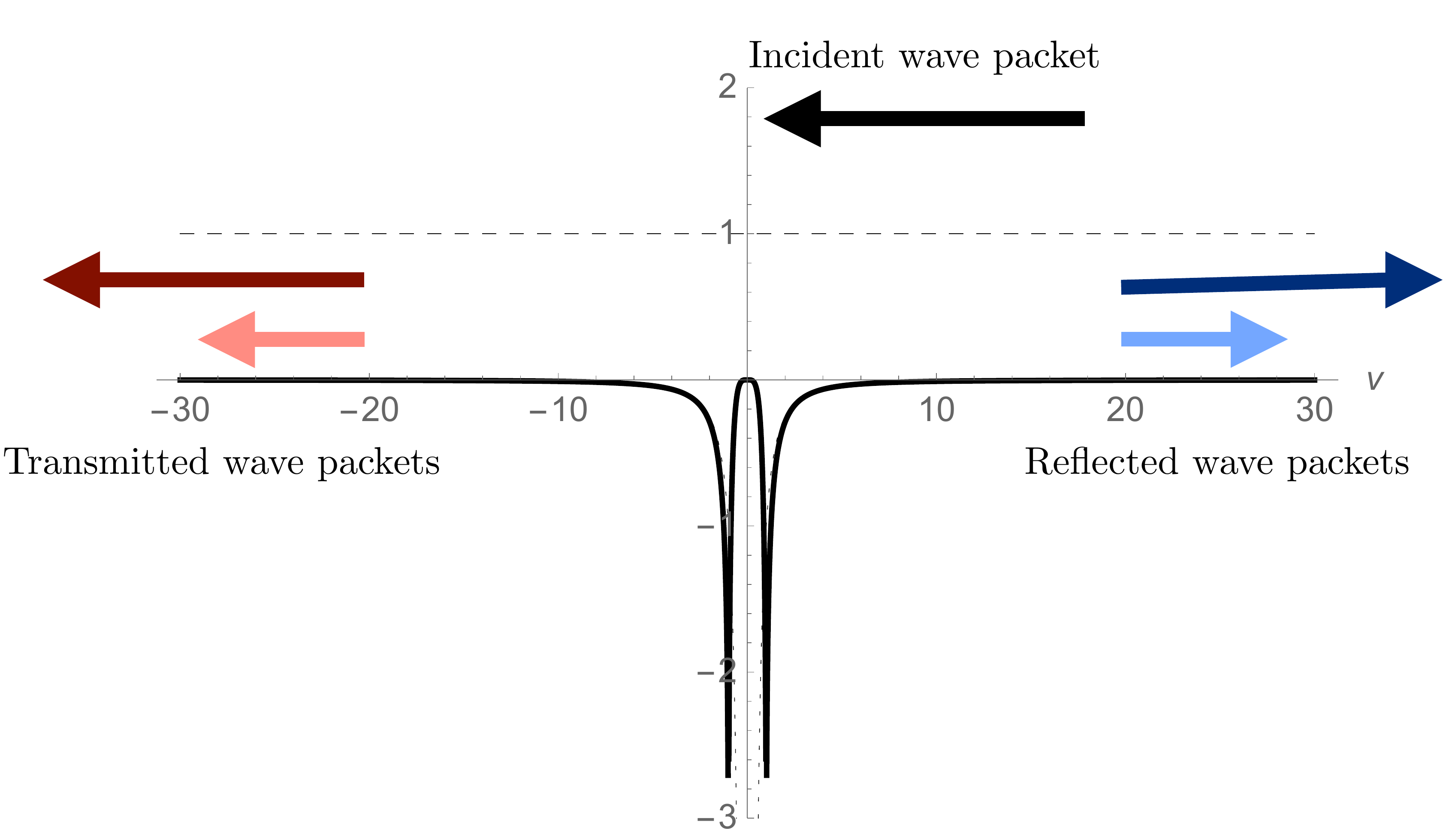} 
\end{array} 
\)} 
\caption{In the traditional polymerization $f^2(x)=(\sin(x))^2$ (dispersion relations on the left) new solutions appear. 
On the right panel we show the different scattering channels in a matter coupling with a massless scalar that produces (for a given $p_{\phi}$ eigenstate)
an effective potential regularized by inverse volume corrections (shown in black). 
The universe bounces and tunnels in new channels with different asymptotic Hubble rates for a given cosmological constant. If the $v\to -v$ symmetry is imposed (as customarily in the specialized literature) the degeneracy remains and the universe only bounces yet into the quantum  superposition of two semiclassical solutions. The results of an analytically solvable model are shown in \eqref{QS}.
} \label{diff2}  
\end{figure} 

\subsection{Dynamical consequences when matter couplings are included}

 Here we show how the inclusion of matter couplings has the generic effect of producing `diffusion' into the various energy eigensectors which would not be present in the Schroedinger quantization. The additional energy eigenvalues of the pure gravity model introduced by the choice of polymerization play an important dynamical role. We will see that a universe starting in the large volume limit in one asymptotically De Sitter state---with a given cosmological constant (energy) and a given Hubble rate $b_0$---will `scatter' through the big bang into  a superposition of the various eigenstates of the same asymptotic energy. In this way, the quantum dynamics of the bounce is way more complicated than hinted by the effective equation approach that will be constructed in the later part of the paper. This is a simple instance of the physical expectation embodied in the statement that {\em anything that can happen happens in quantum mechanics. }
It shows in a crystal clear way that, in a background independent approach, the most likely result is that an initially semiclassical state (with a clear spacetime interpretation)
will evolve into a superposition that might not always admit a {\em single spacetime} representation. Forgetting this simple fact about quantum mechanics is one of the most current errors in setting up important questions such as the ones concerning the fate of information in black hole evaporation. Here again we see how, the present models of quantum cosmology, represent a rich and valuable testground for conceptual ideas in spite of their limited quantitative predictive power as far as observable effects are concerned.

One of the simplest models of matter coupling is that of a massless scalar field (i.e. $U(\phi )=0$ in equation \eqref{2dis}). In that case the momentum of the scalar field is conserved and the gravitational dynamics is equivalent to that of a point particle (with kinetic energy $\propto b^2$) moving in a `external attractive potential' that goes like $\propto 1/v^2$ (see equation \eqref{hammm}). The divergence in the potential is regularizaed in loop quantum cosmology using the Thiemann construction that modifies the inverse volume dependence near the big bang at $v=0$. Such modification is illustrated in the Figures \ref{diff1}, \ref{diff2}, \ref{diff3}. Such model is already complicated enough to make analytic statements involved. 

However, the qualitative behaviour that one want to illustrate does not depend on the details of the potential and only on the fact that the coupling with matter breaks the conservation of the variable $b$. This is simply due to the fact that matter couplings break translational invariance in the $v$-axes producing a non trivial dynamics of the Hubble rate (a quite obvious fact from the standard classical perspective based on the Friedmann equations where only in pure DeSitter spacetime the Hubble rate remains constant in the FLRW slicing). Thus, the phenomenon we want to emphasize can be illustrated in a much simpler model where analytic calculations are trivial. An example of such model is the one where the regularized $1/v^2$ potential produced by  the corresponding contribution to the Hamiltonian \eqref{Hamon} of the massless scalar field is replaced by the sum of two Kronecker deltas at $v=0$ and $v=\lambda$ mimicking in some way the two picks in the regularized potential seen in the previous figures. Notice however, that this example is not meant to approximate in any precise sense the massless scalar field case. We are only using it because we can solve it explicitly and because it produces the phenomenology that will be common (at the qualitative level) to any matter coupling. The only essential feature here is its breaking of translational invariance in the $v$-axes.

Concretely, we concentrate on the difference equation
\be
\Psi_e(v-2\lambda)+\Psi_e(v+2\lambda) + (e-2) \Psi_e(v)-\alpha \delta\left(\frac{v}{\lambda},1\right) \Psi_e(v)-\alpha \delta\left(\frac{v}{\lambda},0\right)\Psi_e(v)=0,
\ee
where $e\equiv 8m\lambda^2 E$ and the delta functions are Kronecker deltas on the lattice $v=\lambda n$ with $n\in \Z$, and $\alpha$ is a coupling constant. This is a simple scattering problem which is resolved via the ansatz
\begin{equation}\label{eq151}
\Psi_{b_1}(v)  =
\begin{cases}
e^{- i  {b_1(e)}  v} + R_1(e) \, e^{i {b_1} v} + R_2(e) \, e^{i {b_2} v} & \text{($v \ge 0$)} \\
T_1(e)\, e^{- i  {b_1(e)}  v}+ T_2(e)\, e^{- i {b_2(e)}  v}& \text{($v \le 0$)},
\end{cases},
\end{equation}
where $b_1(e)>b_2(e)>0$ are the two positive solutions of the dispersion relation---plotted in Figure \ref{diff2}---for the given value of $e$.
The previous discrete Schroedinger equation boils down to 4 independent linear equations from which one determines that reflexion and transmission amplitudes. They are given by 
\ba \label{QS}
&& \!\!\!\!\!\!\!\!\!\! \!\!\!\!\!\!\!\!\!\!  R_1(e)=\frac{\alpha  \left(i \alpha 
   \sqrt{-(e -1) e
   }+\alpha 
   (1-e )-(1-e)
   \left(e -i
   \sqrt{-(e -1) e
   }\right)\right)}{(e -1)
   \left(\alpha ^2+e \right)} \ \ \ \ \ \  R_2(e)=\frac{i \alpha ^2 \sqrt{e
   }}{\sqrt{1-e }
   \left(\alpha ^2+e \right)}\n \\
&&  \!\!\!\!\!\!\!\!\!\! \!\!\!\!\!\!\!\!\!\! T_1(e)=\frac{e }{\alpha ^2+e
   }-\frac{i \alpha  \sqrt{e
   }}{\sqrt{1-e }
   \left(\alpha ^2+e \right)}\ \ \ \ \ \ \ \ \ \ \ \ \ \ \ \ \ \ \ \ \ \ \ \ \ \ \ \ \ \ \ \ \ \ \ \ \ \ \ \ \ \ \ \ \ \ \ \ \ \ \ T_2(e)=\frac{\alpha  \left(1+\frac{i
   \sqrt{e
   }}{\sqrt{1-e }}\right)
   e }{\alpha ^2+e }.
\ea
There are two interesting limits of the previous result that will be relevant for the discussion 
in Section \ref{tunnel}: the reflection amplitudes vanish in the limit $\alpha\to 0$ where the bounce is completely suppressed, while the transmission amplitudes vanish in the hard-scattering limit $\alpha\to \infty$. Here we have used a simplistic model where explicit calculations can be done. As mentioned before, the qualitative features present in this model remain in the realistic case (this is confirmed by numerical simulations that we have omitted form the paper for simplicity).

\section{The modified cosmological effective equations}

Let us start from the unimodular Hamiltonian constraint where \eqref{Hamon} is equated to some energy value that plays the role of the cosmological constant, namely 
\be  \label{coco}
C\equiv {\cal H}-V_0\frac{\Lambda}{8\pi G}=\frac{V_0}{2\pi G \gamma^2}\left(\frac{3f(\lambda b)^2}{4\lambda^2}-\frac{p_\phi^2}{16 \pi G v^2}-2\pi G \gamma^2 \, U(\phi)-\frac{1}{4} \Lambda\right)\approx 0,
\ee
Let us study the evolution of the volume variable \be v=\frac{V_0 a^3}{4\pi G\gamma}\ee
\begin{equation}\label{eq:hr_02}
\frac{d\braket{v}}{ds}=-i\braket{\left[v, {\cal H} \right]}\approx \frac{\partial \braket{{\cal H}}}{\partial b} =\frac{3 V_0}{4\pi G\gamma^2 \lambda}  f'( \lambda b) f( \lambda b) 
\end{equation}
where we have used the results of Appendix \ref{appA} in the derivation of the effective equations, and 
prime denotes derivatives with respect to $ \lambda b$, and s denotes unimodular time given in terms of co-moving (cosmic) time $\tau$ by
\ba \label{titi} ds&=&-|a|^3 d\tau\n \\
&=&-\frac{4\pi G\gamma}{V_0} |v| d\tau. \ea 
Indeed, the previous equation gives us an expression for $\dot a/a$, namely
\be\label{HH}
\frac{\dot a}{ a}= -\frac{1}{\gamma \lambda} f'(  \lambda b) f( \lambda b) 
\ee
From now on we denote $\braket{v}$ simply $v$. Using the standard definition of the Hubble rate $H\equiv \dot a/a$ we can write
\begin{equation}\label{eq:hr_03}
    H^2= \frac{1}{\gamma^{2} \lambda^{2}} f'(  \lambda b)^{2} f( \lambda b)^{2}.
\end{equation}
The constraint \eqref{coco} can be rewritten as
\be\label{contri}
\boxed{C=\frac{3V_0}{8\pi G\gamma^2\lambda^2}\left(f(\lambda b)^2-\frac{\rho+\rho_\Lambda}{\overline{\rho}}\right)\approx 0}
\ee
 where $\rho_\Lambda\equiv \Lambda/(8\pi G)$ 
\be \bar{\rho} = \frac{3}{8\pi G \gamma^{2} \lambda^{2}},\ee
and
\ba\label{densi}
\rho&=&\frac{p_\phi^2}{32\pi^2 G^2 \gamma^2 v^2} +U(\phi)+\frac{\Lambda}{8\pi G}\n \\ 
&=&\frac{\dot\phi^2}{2}+U(\phi)+\frac{\Lambda}{8\pi G}
\ea
is the standard energy density. This implies \be f( \lambda b) = \sqrt{\left( \rho/ \bar{\rho} \right)}.\ee It is also convenient to introduce the pressure
\ba\label{press}
P&=&\frac{p_\phi^2}{32\pi^2 G^2 \gamma^2 v^2} -U(\phi)-\frac{\Lambda}{8\pi G}\n \\ 
&=&\frac{\dot\phi^2}{2}-U(\phi)-\frac{\Lambda}{8\pi G}.
\ea
We arrive thus to the modified Friedman equation
\begin{equation}\label{fried}
        \boxed{H^{2}=  \frac{8 \pi G\rho }{3}  \left[f ' \left( f^{-1} \left( \sqrt{ \rho/ \bar{\rho} } \right) \right)\right]^{2}.}
\end{equation}
We can now look at the evolution equation of the energy density \eqref{densi}
\ba\label{eq:conservation_law}
    \frac{d\braket{\rho}}{ds}&=&-i\braket{[\rho, {\cal H}]}
    \n \\ &=&-i \frac{3 V_0}{8 \pi G\gamma^2\lambda^2}\braket{[\rho, f(\lambda b)^2 ]} = \frac{3}{8 \pi G\gamma^2\lambda^2}\frac{d\braket{f(\lambda b)^2} }{ds}\ea 
where we have used that the Hamiltonian is ${\cal H}=3V_0 f(\lambda b)^2/(8\pi G\gamma^2\lambda^2)-V_0\rho$. Now using Remark \ref{remyy} we get
\ba
    \frac{d\braket{\rho}}{ds}&=& -\frac{3}{16 \pi G\gamma^2\lambda^2} \frac{\braket{p_\phi^2}}{(4\pi G \gamma^2) v^3} 4 \lambda f'(  \lambda b) f( \lambda b) \n \\
    &=&  \frac{3}{ 16 \pi G\gamma^2\lambda^2} \frac{4 \braket{p_\phi^2}}{(4\pi G \gamma^2) v^3} \lambda^2\gamma H.
    \ea
An important corollary of the previous algebra (or simply from Remark \ref{remyy}) is that 
\be\label{bdoty}
 \dot{b} =
-4\pi G \gamma  (\rho + P)  \frac{|v|}{v},
\ee
which follows directly from Remark \ref{remyy}, the definition of $\rho$ and $P$, and the sign comes from the relationship \eqref{titi} between comoving time $\tau$ and unimodular time $s$.    
Now using \eqref{titi} we get the continuity equation
\be\label{conti}
\boxed{\dot \rho+3H (\rho+P)=0}
\ee  
where the quantities in the equation are to be taken as expectation values.
It is now a simple exercise to show that from \eqref{conti} and \eqref{fried} the following \textit{modified} Raychaudhuri equation
follows
\begin{equation}\label{eq:hr_03_b}
    \dot{H} = -4 \pi G \left( \rho + P \right) \left[ f ' \left( f^{-1} \left( \sqrt{ \rho/ \bar{\rho} } \right) \right)^{2} + f''\left( f^{-1} \left( \sqrt{ \rho/ \bar{\rho} } \right) \right) f\left( f^{-1} \left( \sqrt{ \rho/ \bar{\rho} } \right) \right) \right]. 
\end{equation}
This concludes the derivation of the effective cosmological equations for arbitrary regularizations of the Hamiltonian encoded in the arbitrary function $f(\lambda b)$.
We see that in regions where the latter behaves linearly as in \eqref{naive}, one recovers the standard classical Eintein's equations in the cosmological context. However, and this is the key point of this paper, deviations from Einstein's equations can be introduced by `tuning' the function $f(\lambda b)$. Such modifications, as we will see, do have important physical consequences and thus make the large number of Fourier coefficients in $f(\lambda b)$ relevant ambiguity parameters compromising the use of these models for physical predictions.

\section{The landscape of polymerized models of quantum cosmology}

In this section we analyse the generic implications of the effective dynamical equations. We will assume their validity through the region corresponding to the would-be-singularity of classical cosmology where the scale factor $a$ approaches zero. Even when for certain suitable initial semiclassical states for contracting universes this approximation might hold true in some cases (for instance for suitable bouncing solutions) we know from our analysis of Section \ref{qqq} that the state of the universe branches off into other solutions that go right through the $a=0$ regime. In these other branches the effective dynamical equations break down unless one considers a rather artificial regularization of the inverse volume corrections. This is why such solutions can only be understood in full generality using the quantum theory. Due to this behaviour we call these solutions \emph{tunneling solutions}.

\subsection{Bouncing branches}\label{bouncingg}

We will first study the bouncing solutions of the effective equations that are usually described in the LQC literature, analysing the generic effect of the choice of the polymerization 
function $f(\lambda b)$.
We recall equation \eqref{HH}
\be\label{aa}
 \frac{\dot a}{a}= -\frac{1} {\kappa \gamma} f'(\lambda b) f(\lambda b) \quad,
\ee
where, $f'(\lambda b)$ denotes the derivative with respect to $\lambda b$. We also need the modified Raychaudhuri equation  \eqref{eq:hr_03_b} which can be written in the form
\begin{equation}\label{bb}
    3 \frac{\ddot{a}} {a} = -{4 \pi G} \lpr (\rho + 3P) f'^2 + 3 (\rho + P) f'' f \rpr + \Lambda f'^2 \quad.
\end{equation}
From  \eqref{bdoty} and assuming the validity of the {\em null energy condition} (NEC), $\rho + P \ge 0$, we can determine the direction of change of $b$ depending on the sign of the volume of the universe. This greatly simplifies the analysis of the landscape dynamics. NEC are violated due to quantum gravity effects when inverse volume corrections in the matter coupling are taken into account. However, this is not relevant for the bouncing branches for states such that the effective equations are valid as the bounce prevents $v$ from reaching the regions where $NEC$ are violated (see below Section \ref{nec}).

\begin{figure}
\begin{center}
\includegraphics[width=0.6\textwidth]{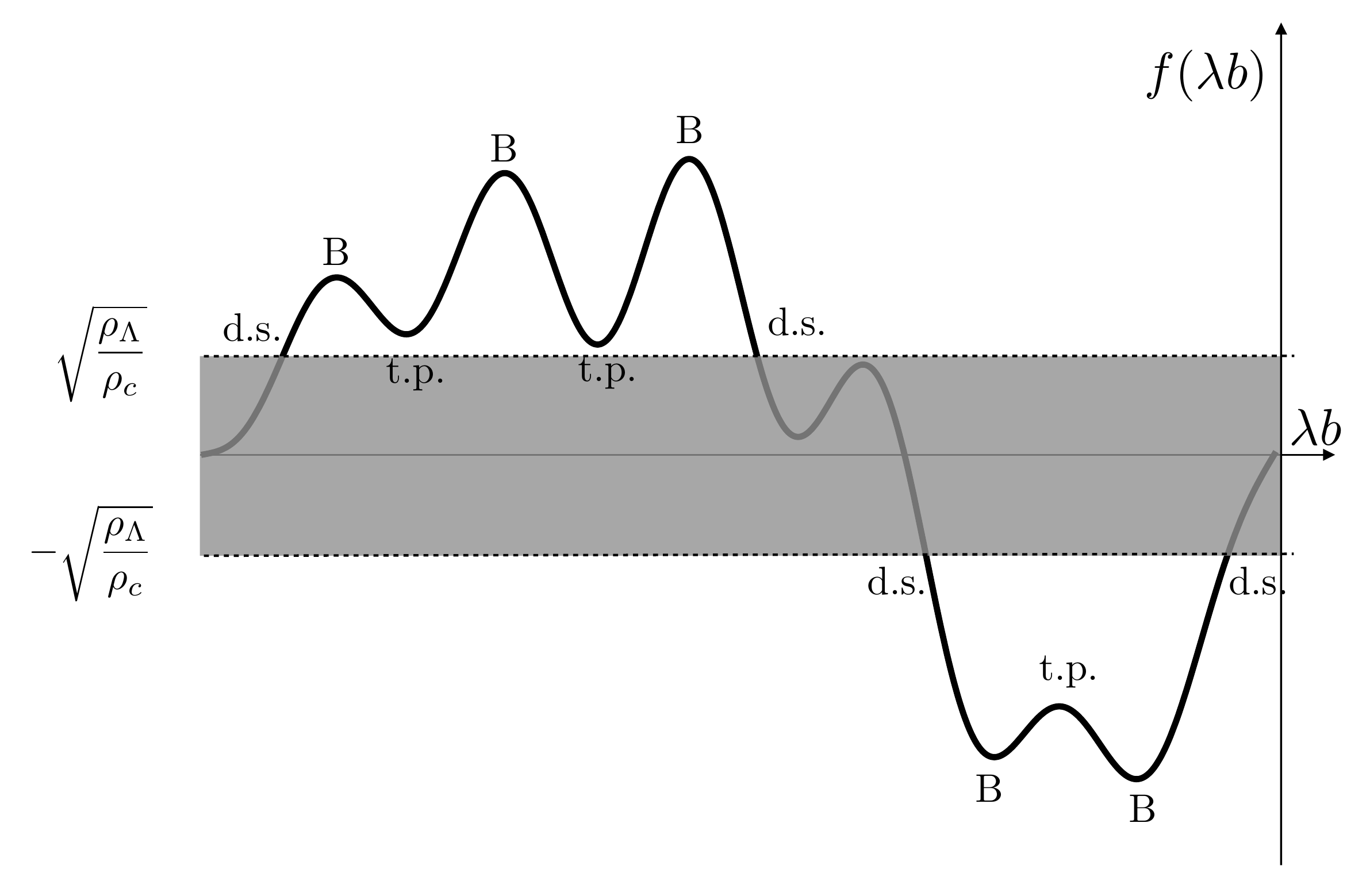}
\caption{Illustration of the dynamical features brought by the use of a generic $f(\lambda b)$  when assuming that $\Lambda \ge 0$. If we require $\rho\ge0$, we have a bound on $f$ given by $f^2\ge \rho_\Lambda / \bar\rho$. Close to $\lambda b_0$ with $f^2(\lambda b_0)= \rho_\Lambda / \bar\rho$ the function must be well approximated by $f(\lambda b)\approx \pm \lambda (b-b_0)$ in order to recover general relativity at low regular matter densities. We denote bounce points with B, turning points with $\rm t. p. $, and fixed points where the universe becomes asymptotically De Sitter with $\rm d.s.$}
\label{fig:bounce_pos}
\end{center}
\end{figure}

Critical points in the function $f(\lambda b)$ correspond to two possibilities: bounces (minimum volume configurations where the universe stops contracting and starts expanding) and turning points (maximum volume configurations where the volume of the universe stops increasing and starts decreasing). Such situations are identified by the condition $\dot{a} = 0$, which, from \eqref{aa}, arises when $f = 0$ and $f' = 0$. We will study first the case $f' = 0$. In order to understand if we are at the presence of a bounce or a turning point, we have to study the sign of the second derivative of the volume (a bounce occurs for $\ddot v > 0$, a turning point for  $\ddot v <0$). Evaluating \eqref{bb} at points where $f^\prime=0$ we get
\begin{equation}
    \left. \frac{\ddot a} {a} \right|_{f'=0} = - 4\pi G (\rho + P) f'' f \quad.
\end{equation}
Assuming that the NEC is valid, local maxima of the function $f$ represent a bounce when $f > 0$ as well as for local minima of $f$ when $f < 0$. Conversely, we have turning points at local minima of $f$ for $f > 0$, and at local maxima of $f$ for $f < 0$. What about those points where $f=0$?
If the cosmological constant is positive, then the from of the Hamiltonian imposes that 
\be
f^2(\lambda b)\ge \sqrt{\frac{\rho_\Lambda}{\rho_c}},
\ee
so that these points are inside a `classically forbidden' region but they can be reached by setting $\Lambda=0$. In this case they become a special case of points where $f(\lambda b)^2 \rho_c=\rho_\Lambda$. In general there are De Sitter asymptotic configurations (Minkowski being a limiting case) where the contribution of other forms of matter to $\rho$ and $P$ vanish. All this is illustrated in Figure \ref{fig:bounce_pos}. Assuming that the universe is in the $v>0$ branch, then equation \eqref{bdoty} implies that  $\dot b\le 0$ as long as the NEC hold with $\dot b=0$ at the De Sitter configurations where $\rho+P=0$. These configurations are fixed points of the flow of $b$ (recall that $b$ is simply related to the Hubble rate according to the classical analysis from where we started).  

With all this information we can interpret the situation described in Figure \ref{fig:bounce_pos} qualitatively by observing that it represents two distinct possible histories for the universe. The first starts on the  classically allowed region on the right in an (asymptotically) De Sitter state defined by the furthest intersection of $f(\lambda b)$ with $\sqrt{\rho_\Lambda/\rho_c}$ to the right. The universe contracts and goes out of the purely De Sitter state entering a phase where other forms of matter start  playing a dynamical role. At the first minimum the universe bounces for the first time and starts expanding. The expansion continues until the universe gets to the first maximum (from right to left) where it starts contracting again until the second minimum is reached and a last bounce leads to an expanding universe that expands forever towards a final asymptotic De Sitter state.  A second sequence of events can be described in a similar fashion for the evolution along the classically allowed region on the left of  Figure \ref{fig:bounce_pos}.

Note that the initial and final asymptotic De Sitter phases are described by different Hubble rates are the former is modulated by the value of the $f^{\prime 2}$ at the asymptotic points according to \eqref{fried}. One could introduce an effective cosmological constant at such De Sitter fixed points 
\be\label{ddss}
\Lambda_{\rm ds-fp}\equiv \Lambda f^{\prime 2}.
\ee  
Notice that these fixed point correspond to low energy regions where the density of matter (other than the cosmological constant) goes to zero.
More generally, one can expand the modified Friedman  \eqref{fried}
 around an arbitrary density $\rho_0$ and write  \be
\frac{\dot a^2}{a^2}=\frac{8\pi G_{\rm eff}}{3} \rho+\frac{\Lambda_{\rm eff}}{3}+\sO\left(\frac{(\rho-\rho_0)^2}{\rho^2_c}\right)
\ee
A simple calculation gives
\be
G_{\rm eff}=G (f^{\prime 2}+f^{\prime \prime} f),
\ee
and
\be
\Lambda_{\rm eff}=(f^{\prime 2}+f^{\prime \prime} f)\Lambda- 8\pi G \rho_0 f f^{\prime\prime}
\ee
For $\rho_0=\Lambda/(8\pi G)$ we recover the De Sitter fixed point value \eqref{ddss}.
Notice, that as we approach a bouncing point $G_{\rm eff}$ becomes negative turning gravity repulsive.
The same functional dependence of $G_{\rm eff}$ is responsible for the effect interpreted as a change of signature in \cite{Bojowald:2015gra}.

Finally, in the case of $\Lambda < 0$ there is a non trivial lower bound for $\rho$  given by $\rho_{\text{min}} = - \rho_\Lambda $ corresponding to points where $f=0$. 
In the point particle analogy these are to turning points of a bound state where the kinetic energy vanishes, here $f(\lambda b)=0$. For the universe these are turning points where the universe achieves minimal regular matter density before recollapsing in to a denser regime. The other qualitative features at critical points remain the same as in the previous discussion.

\subsection{Tunnelling branches (in the case of a massless scalar field)}\label{tunnel}

In Section \ref{qqq} we have shown that the quantum theory predicts that, in addition to the traditional bounce evoked in the previous discussion and advocated in the LQC literature at large, the universe can also tunnel across the singularity into an expanding phase. The bounce is something that is clearly captured by the effective equations. Can tunnelling also become apparent from these equations?  It is possible to see this if one considers for a moment inverse volume corrections in the matter coupling described in Section \ref{invv}.
This perspective will turn out not to be important by the end of the discussion in this section; however, it gives a concrete classical classical picture of the process we have in mind. We will see that for this classical picture to hold one has to push to the extreme non-Planckain region the inverse volume corrections in the inverse volume regularization. However, the dynamical channel remains open in the fundamental quantum theory where semiclassical descriptions are only a good interpretational tool away from the big bang.

For simplicity we concentrate on the case of a massless scalar field matter model. Assuming that we are in an eigenstate of the momentum $\pi_\phi$ (conserved in this case) we have already observed that its contribution to the Hamiltonian \ref{Hamon} can be seen as an effective potential in the analogy with a non relativistic particle parametrization of Section \ref{uuu}. However, this contribution is now everywhere finite as the $1/v^2$ classical behaviour is regularized by the Thiemann trick. This means that there must exist solutions of the effective equations where the the universe evolves right through $v=0$ (or scale factor $a=0$) into the $v<0$ without experiencing the bounce produced by the kinetic term when the variable $b$ reaches the suitable critical points of $f(\lambda b)$ described above. For this to happen the universe must scatter thought the big bang singularity `softly' in the sense that the variable $b$ must not grow up to one of the critical points of $f(\lambda b)$. This happens when the universe---interpreted as the point particle---rolls down the potential  
\be V(v)\equiv-p_\phi^2 \left[\frac 1{v^2}\right]_{\rm reg}\ee in a way such that its `kinetic energy' does not grow beyond the bound
\be\label{ofo}
\frac{p_\phi^2}{4 \pi G\gamma^2}{\rm max} \left[\frac 1{v^2}\right]_{\rm reg} \le \frac{3}{\gamma^2\lambda^2} f^2(\lambda b_c)-\Lambda=\frac{3}{\gamma^2\lambda^2} (f^2(\lambda b_c)-f^2(\lambda b_\infty)).
\ee
We observe that, as the regularized potential is bounded, for $|p_\phi|$  sufficiently low the universe experiences (according to the effective equations) a soft-transition from $v>0$ to $v<0$ in finite unimodular time $\Delta s$ instead of a bounce. The scale factor crosses $a=0$; however, there is no singularity as one can easily check from the effective equations \eqref{aa} and \eqref{bb} and the fact that both $P$ and $\rho$ vanish there. Indeed, the universe goes through a De Sitter phase where $\Lambda$ dominates. Even when $a=0$ is reached in finite unimodular time $s$, the would-be-singularity $a=0$ is reached at infinite comoving time $\tau$. If valid, the effective equations predict an infinite number of e-folds of inflation at around the soft-transmission from $v>0$ to $v<0$. Even when here there is only one such transitions, the scenario resembles in spirit the eon-transition of {\em conformal cyclic cosmology} proposed by Penrose \cite{Penrose:2010zz}. However, these conclusions are not really correct for inverse volume regularizations that modify the matter coupling from the expected classical behaviour at around the Planck scale only. 

Concretely, if we take the standard inverse volume correction based on $[1/\sqrt{v}]_1$ (recall equation \ref{decho}) for which 
\be
{\rm max} \left[\frac 1{v^2}\right]_{\rm reg}=\frac{4}{\lambda^2\hbar }
\ee 
\be
{ p_\phi^2} \le 3 \pi \ell_p^2 \left(f^2(\lambda b_c)-f^2(\lambda b_\infty)\right).
\ee
However, we see that in order to trust the previous conclusions the effective equations would have to be valid in the description of the universe from $v=\lambda$ to $v=-\lambda$. If $\lambda$ is taken to be of the order of the Planck scale then it is clear that the details of the dynamics evoked above (De Sitter inflation for an unlimited number of e-folds) does not survive in the fundamental description where the variable $v$ jumps on discrete values of the order of $\lambda$. However, the conclusion that the transmission channel exists in addition to the well known bouncing channel remains.  A precise analysis of such transitions would require using a fully quantum treatment which is of course possible. 

In this respect it is interesting to revisit the results of Section \ref{qqq} under the light of the present discussion. Notice that, qualitatively speaking, the parameter $\alpha$ regulating the strength of the toy-model potential in \eqref{QS} is the analog of $p_{\phi}^2$ here. We observe that, even when always non vanishing,  the transmission amplitudes go to zero in the limit $\alpha\to \infty$ and only the bouncing channels remain available. 
Consideration of the quantum theory uncovers a feature that we have evoked previously.
Indeed, the criterion for soft bounce \eqref{ofo} looses its quantitative relevance and we realize that even if one considers un-bounded regularizations such as the one proposed in \cite{Wilson-Ewing:2012dcf, Singh:2013ava} there will be a component of wave function in the transmission sector in addition to the bouncing sector for suitable initial states that probe the potential on sufficiently soft points of the potential. More precisely, consider the regularization where
\be
\left[\frac{1}{v^2}\right]_{\rm reg}=\frac{1}{v^2}  \ \ \ \forall \ \ \ v\not=0, \ \ \ {\rm while} \ \ \  \left[\frac{1}{0^2}\right]_{\rm reg}=0. 
\ee
Consider a semiclassical state defined on a lattice of $v=n \lambda$ with $n\in \Z$, i.e. a superposition of volume eigen-states that will evolve on this lattice in such a way that the potential will be probed only on such lattice points. The criterion \eqref{ofo} can be written in this case as
\be
{ p_\phi^2} \le p^2_{\rm C}\equiv 12 \pi \ell_p^2 \left(f^2(\lambda b_c)-f^2(\lambda b_\infty)\right).
\ee
Now this cannot be a sharp bound because its construction relies on the effective dynamics. However, it remains an order of magnitude criterion in the sense that  as $p_\phi^2$ becomes smaller and  $p_\phi^2\ll p^2_{\rm C}$ the transmission probability is expected to dominate while the bouncing probability will become smaller and vice versa.  
Indeed a more direct dimensional analysis argument is perhaps clearer. Assuming the change in the function $f(\lambda b)$ in the region of interest is order unity (which is about right for a continuous function unless one would dramatically tune $f(\lambda b)$) then the criterion of softness is very simple and boils down to the condition that\footnote{It is interesting to notice that if we assume that the Universe is described by an massless scalar field before the onset of inflation one can estimate $p_{\phi}$ as follows. Using equation \eqref{densi} we have 
\begin{equation}
    \rho \sim \frac{p_{\phi}^2}{V_{0}^2 a^6}.
\end{equation}
If we assume that the density of the onset of inflation is $\sim 10^{-5} m_{\mathrm{p}}^4 $---as is the case, for instance, in power law inflationary model \cite{Weinberg:2008zzc}---and we take the physical volume of the fiducial set at the onset of inflation to be $\sim 10^2 m_{\mathrm{p}}^{-3}$ we obtain that $p_{\phi}^2 \sim 10^{-1} m_{\mathrm{p}}^{-2} \sim 10^{-1} \ell_{\mathrm{p}}^{2}$. Note that that a fiducial cell with physical volume of $10^2$ Planck volumes will inflate to a size much larger than the observable universe today. For a discussion of the role of $V_0$ in quantum fluctuations see \cite{Rovelli:2013zaa}, this of course adds and additional dimension to the ambiguity discussion.}
\be\label{tututu}
p_{\phi}^2\lesssim \ell_p^2.
\ee

\subsection{Violation of the NEC due to inverse volume corrections}\label{nec}

\begin{figure}[h]  
\centerline{\hspace{0.5cm} \(
 \begin{array}{c}
\includegraphics[width=8cm]{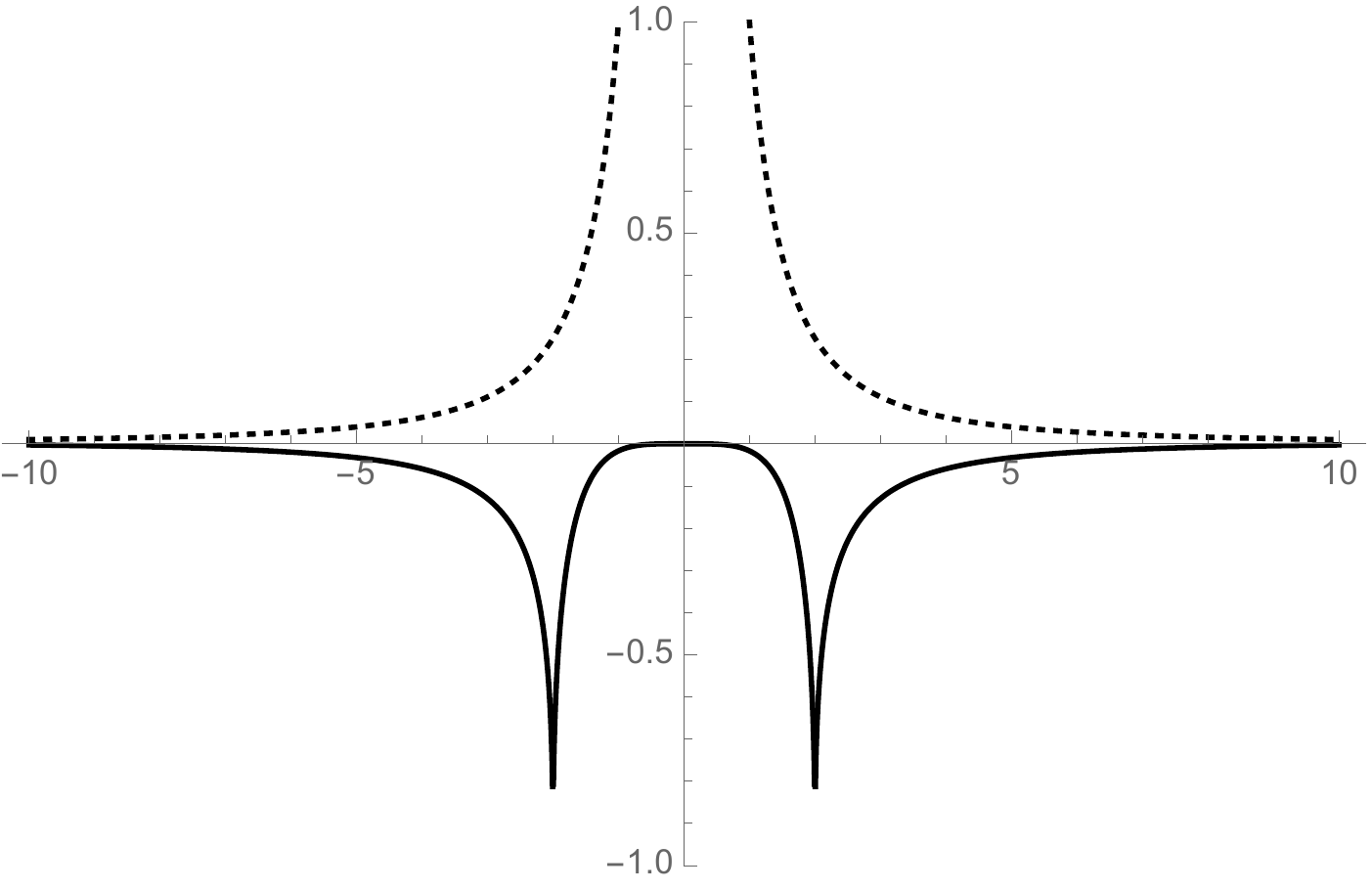} 
\end{array} \ \ \ \ \ \ \ \ \ \ \ \ \ \ 
 \begin{array}{c}
\includegraphics[width=7cm]{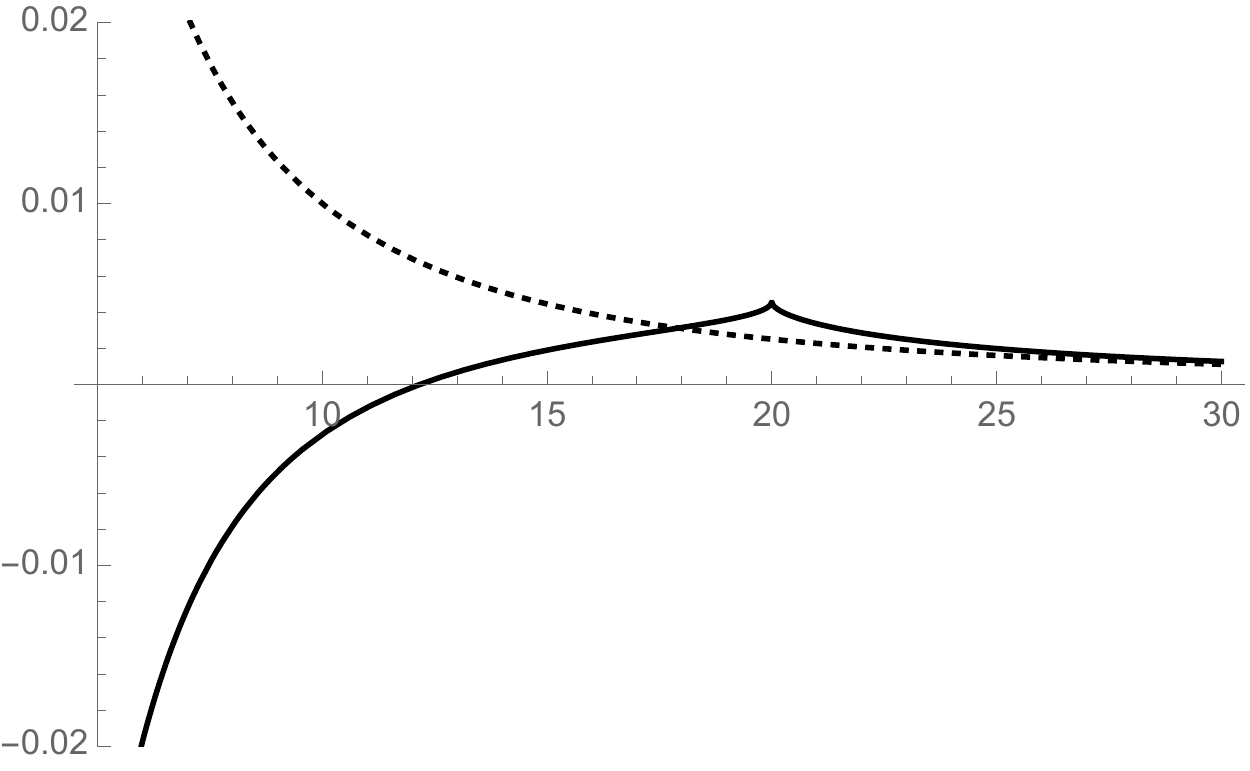} 
\end{array} 
\)} 
\caption{Regularization of the function $1/v^2$: the classical function corresponds to the dotted line. The regularization shown here violates the positivity in a range around $v=0$ and coincides in the IR with the classical function as the plot on the right illustrates. This leads from equation \eqref{jijiji} to violations of the NEC that when considered in the Hamiltonian produce a different type of bounce for suitable initial conditions. The plot represents the regularization given in \eqref{reputamadre}.} \label{NEC}  
\end{figure}

The NEC requires that $T_{ab} k^ak^b\ge 0$ for any future directed null vector. In our cosmological setting any matter coupling can be considered as a perfect fluid as demanded by isotropy, and thus the NEC reduces to the statement that $\rho+P\ge 0$. For a scalar field model of the type considered here (and independently of the self-interaction potential) this condition is classically given by 
\be\label{jijiji}
\rho+P=\frac{p_\phi^2}{16\pi^2 G^2 \gamma^2 v^2},
\ee
which satisfies the NEC trivially.
In the quantum theory the NEC can be violated by the inverse volume corrections introduced by a regularization, for instance of the class \eqref{decho}.
As an example we plot the regularization 
\be\label{reputamadre}
\left[\frac{1}{v^2}\right]_{\rm reg}=\left( 2\left[\frac{1}{\sqrt{v}}\right]_{20}-\left[\frac{1}{\sqrt{v}}\right]_2\right)\left(\left[\frac{1}{\sqrt{v}}\right]_2\right)^3
\ee 
in Figure \ref{NEC}, where one observes that the NEC are violated near the big bang. Such possibility (which is again directly related to the ambiguities of the polymer quantization) has a strong dynamical effect. In the special case of a massless scalar field the previous effect implies also a violation of the weak and the strong energy conditions in the matter sector. When translated into the non relativistic particle analogy of Section \ref{uuu}, one observes that the effective potential in the Hamiltonian \eqref{Hamon} is no longer negative definite. This implies that for sufficiently low cosmological constant---and under suitable conditions where the function $f(\lambda b)$ will play a role---the universe might bounce through yet another different channel due to the repulsive potential produced by the negative energy brought about by the regularization before reaching one of the critical points of $f^\prime=0$.  Once again this is possible if the initial conditions for the matter fields are sufficiently soft so that the probability of this new channel is activated before the standard kinetic bounce described in Section \ref{bouncingg} takes place. A simple analysis that evaluates the amount of `kinetic' energy acquired by the universe (in the non-relativistic particle analogy) as it evolves toward the would-be-singuarity shows that the condition is
\be
\left|\frac{p_\phi}{\ell_p}\right| \lesssim \left|\frac{v_c}{\lambda}\right|,
\ee 
where $v_c$ is the value of $v$ that maximizes the regularization $v^{-2}|_{\rm reg}$ before the NEC are violated (explicitly seen around $v=20$ in Figure \ref{NEC} in this particular example). As $v_c$ can be made large by tuning the inverse volume regularization, the present criterion of softness is weaker than the one for tunnelling \eqref{tututu}.

\subsection{Illustrating examples}

\subsubsection{Inflation induced by ambiguity parameters}

In order to illustrate how the ambiguities of loop quantum cosmology can actually affect the physics in a non trivial manner in this section we show 
that the modifications introduced by the function $f(\lambda b)$ can for instance drive inflation for a large number of e-folds in a way that is weakly dependent of the matter content and dynamics and basically due to the dynamical modifications brought by the `holonomy corrections' in $f(\lambda b)$. We will illustrate this with two simple models: first with a model of a universe filled with thermal radiation, second in the case of a model of inflation with a scalar field with potential $U(\phi)=\lambda\phi^4/2$.
The first example shows that one can get many e-folds of inflation without an inflaton, the second contains an scalar field but the inflation will be driven by effects brought by $f(\lambda b)$, as a consequence, the phenomenology observable in the CMB fluctuations can be tuned, as we will show, by judiciously choosing the later function.   
Thus, let us consider a function $f(\lambda b)$ such that
\begin{equation}\label{eq:hr_04}
   f'( \lambda b)^{2} = 
    \begin{cases}
    1\ \ \ \ \ \ \ \ \ \,  \quad b<b_{c} \\
    \frac{3 H_{0}^{2}}{8 \pi G} \frac{1}{\rho[\lambda b]} \quad b>b_{c}.
    \end{cases}
\end{equation}
where  from  \eqref{coco} we have \be {\rho}[\lambda b]=\bar{\rho} f( \lambda b) ^2,\ee and continuity requires  
\be \label{lili}\rho[\lambda b_c]=\frac{8 \pi G}{3 H_{0}^{2}}\equiv \rho_c ,\ee  with $\rho_c$ a critical density depending on the choice of $H_0$.
A solution of the differential equation \eqref{eq:hr_04} is given by 
\begin{equation}\label{eq:hr_04}
   f( \lambda b) = 
    \begin{cases}\ 
    \lambda b \ \ \ \ \ \ \ \ \  \ \  \ \  \ \  \ \  \ \  \ \  \ \  \ \  \ \  \ \  \ \  \ \  \ \   \,  b<b_{c} \\
    \sqrt{2 \gamma \lambda^{2} H_{0} (b - b_{c}) + \lambda^{2} b^{2}_{c}} \ \ \  \ \  \ \  b >b_{c}.
    \end{cases}
\end{equation}
This choice of $f(\lambda b)$ produces the standard Frieman equation for $\rho\le \rho_c$ while it produces a Friedman equation with constant 
Hubble rate $H_0$ (De Sitter inflation) for $\rho\ge \rho_c$ regardless of the equations of state of matter! The only thing that the matter equations of state will control (via \eqref{conti}) is for how long the universe will remain in the inflationary phase. We will construct two explicit examples in what follows. 

\subsubsection{A pure radiation inflationary model}

The first model consists of a universe filled with radiation $\rho=3 P$. In this case one has from \eqref{conti} that 
\be
\rho=\rho_{\rm in} \frac{a_{\rm in}^4}{a^4}. 
\ee
Assuming that the initial value of $\rho_{\rm in}=m_p^4$ (Planck density) and setting $\rho_c=10^{-68} \rho_{\rm in}$ in \eqref{lili} to the electro-weak transition density\footnote{The electro-weak transition energy scale is $E_{\rm ew}\approx 100$ GeV which corresponds to $E_{\rm ew}\approx 10^{-17} m_p$. } we see that inflation can be sustained as long as  
\be
 \frac{a_{\rm in}^4}{a^4}\ge 10^{-68},
\ee
in other works for a number of e-folds of about
\be
\sN_{\rm rad}=17 \log(10)\approx 39.
\ee
Using a massless scalar field (which is an often used example) with equation of state $P=\rho$ one has
\be
 \rho=\rho_{\rm in} \frac{a_{\rm in}^6}{a^6},
\ee
and the same $\rho_c$ we get $\sN_{\rm \phi}=\log[10] 68/6\approx 26$ which is still a considerable number. 

\subsection{Inflation with a scalar field}

It is also possible to design a model without an inflaton just using the matter content of the standard model of particle physics 
where inflation is driven by polymer corrections. The model is consistent with the observed fluctuations in the CMB if the usual paradigm where quantum fluctuations of the Higgs (with potential ${U}(\phi) = \frac{\alpha}{2} \phi^{4}$) are responsible for the generation of inhomogeneities is used. More precisely, starting from the Klein Gordon equation for the Higgs zero mode
\begin{equation}\label{eq:hr_12}
    \ddot{\phi} + 3 H \dot{\phi} +2 \alpha \phi^{3}  = 0,
\end{equation}
and using the standard \textit{terminal velocity} approximation ($\frac{\ddot{\phi}}{H \dot{\phi}}\ll 1$), the solution of \eqref{eq:hr_12} is given by
\begin{equation}\label{eq:hr_13}
    \phi(t)= \frac{\phi_{0}}{\sqrt{1+ \frac{4 }{3 }\alpha \frac{\phi_{0}^{2}}{H_{0}^{2}}  H_{0}t}} \quad\text{ or }\quad \phi(\sN) = \frac{\phi_{0}}{\sqrt{1+ \frac{4 }{3 }\alpha \frac{\phi_{0}^{2}}{H_{0}^{2}}  \sN}}, 
\end{equation}
where we introduced the number of e-folds $\sN=\log(a)\approx H_0 t$.
Now using that $\rho = \frac{\dot{\phi}^{2}}{2} + \frac{\alpha}{2} \phi^{4}$ and equation \eqref{eq:hr_13} we obtain
\begin{equation}\label{eq:hr_16}
    \rho(\sN) = \frac{\alpha}{2}  \frac{\phi_{0}^{4}}{\left[ 1+ \frac{4 }{3 }\alpha \frac{\phi_{0}^{2}}{H_{0}^{2}}  \sN \right]^{2}} \left( \frac{2\alpha}{3} \frac{\frac{\phi_{0}^{2}}{H_0^2}}{\left[ 1+ \frac{4 }{3 }\alpha \frac{\phi_{0}^{2}}{H_{0}^{2}}  \sN \right]} + 1\right), 
\end{equation}
where $\phi_0$ is the initial value of the scalar field. Let us assume that we want \be \sN=50\ee then previous expression must satisfy the condition \eqref{lili}. 
Introducing the variables $y\equiv m_p/\phi_0$ and $x\equiv \phi_0/H_0$ we can write it as
\be
\frac{3y^2}{8\pi}=\frac{\alpha}{2}  \frac{x^2}{\left[ 1+ \frac{200 \alpha}{3} x^2   \right]^{2}} \left( \frac{2\alpha}{3} \frac{x^2 }{\left[ 1+ \frac{200 \alpha}{3} x^2   \right]} + 1\right),
\ee
which imposes some algebraic restrictions on the initial value $\phi_0$ and the Hubble rate $H_0$. We can solve this numerically. For instance we find a solution 
$\phi_0\approx 10 \, m_p$ and $H_0\approx m_p$ if we choose $\alpha=10^{-3}$ (for such small coupling we could solve the previous constraint analytically neglecting the subleading corrections in $\alpha$ but this is not really important here as we only want to exhibit an example). At the end of inflation $\phi_{\rm end}\approx 9.7 \, m_p$ and the reheating phase could start as in usual approaches such as those of chaotic inflation. This example is perhaps more suitable for our point as here the densities remain Planckian all the way during the inflationary phase so that our deviations from general relativity can be more safely attributed to `quantum gravity' effects. In contrast with the previous example where densities went down to almost standard particle physics densities during the anomalous expansion era. One could investigate the mechanism of structure formation. The point of our example is to show the intrinsic discretional nature of these models which precludes the possibility to actually use them for such predictions.  


\section{Conclusions}

We have investigated regularization ambiguities associated with the so-called polymerization process imposed upon us when quantizing cosmological models using the loop quantum cosmology framework.
We showed that quantitative predictability is compromised by the strong dependence on free parameters. However, some qualitative features are robust and independent of the polymerization choice. Among these one finds the well defined quantum evolution across of the big bang which can also be recover at the level of effective dynamical equations valid for suitable semiclassical states.
Thanks to the fact that our quantum dynamics could be explicitly solved, we were able to exhibit  the existence of new channels (tunneling) for the transition across the big bang that are not apparent at the level of the effective dynamical equations.  This was possible due to the use of unimodular quantum cosmology; however, it is easy to understand that these features hold true in the standard formulation. 

The richness of these models should be relevant for the discussion of conceptual and qualitative issues in quantum gravity. For instance in the discussion of questions of unitarity in the context of the black hole  information puzzle (where some initial studies have been performed \cite{Amadei:2019wjp, Amadei:2019ssp}), or in the context of gravitational collapse where the new tuneling modes discussed here could be simplified toy models relevant to investigate the black to white hole transition paradigm of \cite{Haggard:2015iya, Rignon-Bret:2021jch, Martin-Dussaud:2019wqc, Rovelli:2018okm, Bianchi:2018mml}.    
   
\section{Acknowledgements}

We thank useful discussions with A. Barrau, K. Martineau, K. Noui, C. Renevey, and E. Wilson-Ewing.
We acknowledge support of the ID61466 grant from
the John Templeton Foundation, as part of the ``Quantum Information Structure of Spacetime (QISS)'' project
(qiss.fr),  and I$\varphi$U of Aix-Marseille University.

\begin{appendix}

\section{Some properties of Gaussian states in LQC}\label{appA}

The work of Willis and Taveras  \cite{Taveras:2008ke, Willis:2004br} shows that one can approximate the quantum dynamics of loop quantum cosmology 
by effective classical equations when using suitable semiclassical states defined in terms of gaussian states. However, their analysis does not include the type of generalized regularizations studied in this paper. In this section we show that the effective dynamics approximation continues to make sense for arbitrary regularizations of the quantum Hamiltonian as defined in \eqref{Hamon}. 

Let us specialize to a natural choice of semiclassical states (customarily used in the literature studying the issues involved here \cite{Taveras:2008ke, Willis:2004br}). 
For that we chose a state $\ket{\Psi}\in \sH_{lqc}\otimes \sH_\phi$ as a Gaussian state given by
\begin{equation}\label{eq:eff_02}
   	\ket{\Psi} =  \sum_{v_{n} \in \Gamma_{\lambda}} \int_{\mathbb{R}} d p \ \Psi_{v_0,b_0}(v_{n}) \Phi_{p_0,\phi_0}(p)   \ \ket{v_{n}, p}
\end{equation}
where one is using the basis eigenstates of $v$ and $p$ respectively, with the (physical) inner product 
\be\label{iny}
 \braket{{v_{m}, p'}|{v_{n}, p}}=\delta_{n,m} \delta(p,p'),
\ee
and where the wave function \be \Psi_{v_0,b_0}(v) = \sqrt{\frac{\lambda \sigma_b}{\sqrt{\pi}}} e^{- \frac{\sigma_b^2}{2 } \lpr v - v_{0} \rpr^{2} } e^{i b_{0} \lpr v - v_{0} \rpr },\ee is peaked at the geometry phase space point $(v_0,b_0)$, and the wave function \be \Phi_{p_0,\phi_0}(p) = \sqrt{\frac{\sigma_\phi}{\sqrt{\pi}}} e^{- \frac{\sigma^2_\phi}{2} \lpr p - p_0 \rpr^{2} } e^{i \phi_{0} \lpr p  - p_0 \rpr },\ee 
representing a semiclassical state peaked at the point $(\phi_0,p_0)$ of the scalar field phase space. In the previous expressions $\Gamma_{\lambda}$ denotes a regular lattice with lattice spacing $\lambda$ that identifies of the the so-called super selected sectors of the quantum geometry Hilbert space (for a discussion of the nature of such choice see \cite{Amadei:2019wjp, Amadei:2019ssp} and references there in).
%

\subsection{On the equivalence between calculations using the discrete or the continuum inner products}

The following Lemma gives the means to translating expressions involving discrete sums in the loop quantum cosmology inner product to the more familiar continuous integrals of the Schrodinger representation. 
\vskip.3cm
{\prop\label{jijo} For any operator $O(b,p)=\sum_k o_k(p) e^{ib\lambda k}$, and gaussian semiclassical states as in \eqref{eq:eff_02},  one has that
\be
\boxed{\braket{ O(b,p)} \equiv{\braket{\Psi| O(b,p)|\Psi}}  .} \ee}

\vskip.3cm
\noindent{\em Proof:}   By linearity it is enough to prove the previous statement for the operator $e^{ikb\lambda}$ for arbitrary $k$.  One has 
\ba\label{pipo}
{\braket{\Psi|e^{ikb\lambda} |\Psi}} &=&{\frac{\lambda \sigma_b}{ \sqrt{\pi}}} \sum_{n,m}  e^{- \frac{\sigma_b^2}{2} \lpr  \lambda n - v_{0} \rpr^{2} } e^{-i b_{0} \lpr  \lambda n - v_{0} \rpr } e^{- \frac{\sigma_b^2}{2} \lpr  \lambda m- v_{0} \rpr^{2} } e^{i b_{0} \lpr  \lambda m - v_{0} \rpr }  \braket{n|{m-k}}\n \\ 
&=&{\frac{\lambda\sigma_b e^{i 2 b_{0} \lambda k }}{\sqrt{\pi}}} \sum_{m}  e^{- \frac{\sigma_b^2}{2 } \lpr \lambda m- \lambda k - v_{0} \rpr^{2} }  e^{- \frac{\sigma_b^2}{2} \lpr  \lambda m- v_{0} \rpr^{2} } \n \\ 
&=&{\frac{ \lambda \sigma_b e^{i  b_{0} \lambda k }}{\sqrt{\pi}}} e^{- \frac14\sigma_b^2 {\lambda^2 k^2}}\sum_{m}  e^{- \sigma_b^2 {\lpr  \lambda m- v_{0} - \lambda \frac k2\rpr^{2}}} =  {\frac{\sigma_b e^{i  b_{0}\lambda  k }}{\sqrt{\pi}}} e^{-\frac14\sigma_b^2 {\lambda^2 k^2}}\vartheta_3\left[-\frac \pi 2 \left(k+2\frac{v_0}\lambda\right);e^{-\frac{\pi^2}{\lambda^2\sigma_b^2}} \right]\n \\ &=& e^{i  b_{0}\lambda  k } e^{-\frac14\sigma_b^2 \lambda^2 k^2} \left(1+\sO\left(e^{-\frac{\pi^2}{\lambda^2\sigma_b^2}}\right)\right),
\ea
where
\be
\vartheta_3[u;q]\equiv 1+2 \sum_{n=1}^{\infty} q^{n^2} \cos[2n u]
\ee
 in the first line we used the definition of the state \eqref{eq:eff_02} and \eqref{shifty} and then we just rearranged the sums completing squares to arrive at the final result.

{\coloro \label{one} For any operator $O(\lambda b,p)=\sum_k o_k(p) e^{ikb\lambda}$, and gaussian semiclassical states as in \eqref{eq:eff_02},  one has that
\be
\frac{d \braket{\Psi| O(\lambda b,p)\Psi}}{d(\lambda b_0)} = \braket{\frac{dO(\lambda b,p)}{d(\lambda b)}}. \ee}
The proof of the previous statement follows directly from \eqref{pipo} $\square$.

\vskip.3cm
{\coloro For any operator $f(\lambda b)=\sum_k f_k e^{ikb\lambda}$, and gaussian semiclassical states as in \eqref{eq:eff_02},  one has that
\be \braket{f(b)^2}-\braket{f(b)}^2=  2 f'( \lambda b_0)^2 \ \lambda^2\sigma_{b}^2+\sO(\lambda^4\sigma_{b}^4).\ee}

\noindent{\em Proof:} From Remark 1 we have that  \be \braket{f(\lambda b)}= f(\lambda b_0)+ \frac 14 f''(\lambda b_0) \ \lambda^2 \sigma_{b}^2+\sO(\lambda^4 \sigma_{b}^4).\ee
The present statement follows from the previous equation when applied to $O(b)=f^2(b)$ and $O(b)=f(b)$ respectively and replacing the result in the 
expression of the fluctuation
$\square$.

\vskip.3cm
{\coloro The previous two results imply that 
\be
\frac{d \braket{f(\lambda b)^2}}{db_0}=2 \frac{d \braket{f(\lambda b)}}{db_0}\braket{f(\lambda b)}
+\sO(\lambda^2\sigma^2).
\ee}

\subsection{The generating function and the expectation value of operators depending on the volume}
{\prop  For any operator $O(b,p)=\sum_k o_k(p) e^{ib\lambda k}$, and gaussian semiclassical states as in \eqref{eq:eff_02},  one has the generating function on the left
\be\label{ultimaL}
\boxed{\frac{\braket{\Psi| e^{\mathcal{j} (v-v_0)}O(b,p)|\Psi}}{\braket{\Psi|\Psi}} = \sum_{k} o_k(p)  e^{i  b_{0} k } e^{-\frac 14\sigma_b^2 \lambda^2 k^2+\frac{\mathcal{j}^2}{4\sigma_b^2}-\mathcal{j}\lambda \frac k 2} + \left(1+\sO\left(e^{-\frac{\pi^2}{\lambda^2\sigma_b^2}}\right)\right),} \ee
and the generating function on the right
\be\label{ultimaR}
\boxed{\frac{\braket{\Psi| O(b,p) e^{\mathcal{j} (v-v_0)}|\Psi}}{\braket{\Psi|\Psi}} = \sum_{k} o_k(p)  e^{i b_{0} k } e^{-\frac{1}{4}\sigma_b^2 \lambda^2 k^2+\frac{\mathcal{j}^2}{4\sigma_b^2}+\mathcal{j}\lambda \frac k2}+  \left(1+\sO\left(e^{-\frac{\pi^2}{\lambda^2\sigma_b^2}}\right)\right).} \ee}

\vskip.3cm
\noindent{\em Proof:} 
Consider 
\ba
\braket{{\Psi| e^{ikb\lambda} e^{\mathcal{j} (v-v_0)} |\Psi}} &=&{\frac{\lambda \sigma_b}{\sqrt{\pi}}} \sum_{n, m} \ e^{- \frac{\sigma_b^{2}}{2} \lpr\lambda n - v_{0} \rpr^{2} } e^{-i b_{0} \lpr\lambda n - v_{0} \rpr } e^{- \frac{\sigma_b^{2}}{2} \lpr \lambda m- v_{0} \rpr^{2} } e^{i b_{0} \lpr \lambda m - v_{0} \rpr } e^{\mathcal{j} (\lambda m-v_0)}  \braket{n|{m- k}}\n \\ 
&=&  e^{i  b_{0} k } e^{-\frac 14\sigma_b^2 \lambda^2 k^2+\frac{\mathcal{j}^2}{4\sigma_b^2}+\mathcal{j}\lambda \frac k 2}  \left(1+\sO\left(e^{-\frac{\pi^2}{\lambda^2\sigma_b^2}}\right)\right) ,
\ea
where in the second line we completed the square and performed the gaussian integration. 
Equation \eqref{ultimaR} follows from the last line. A similar manipulation gives \eqref{ultimaL} $\square$.


\subsection{Some statements about the truncation of the Fourier expansion}

Given a bounded square integrable function $f(\lambda b)$ of period $2\pi$  we can write it as
\begin{equation}\label{espan}
    f(\lambda b) = \sum_{n \in \mathbb{Z}} a_n e^{i n \lambda b} \quad ,
\end{equation}
with the coefficients $a_n$ given by
\begin{equation}
    a_n = \frac{1}{2 \pi} \int_0^{2 \pi} f(\lambda b) e^{-i n \lambda b} d(\lambda b) \quad.
\end{equation}
which can be bounded \begin{equation}
    |a_n| \le f_{max} \quad.
\end{equation}
where $f_{\rm max} \equiv \max{|f(\lambda b)|} $.
Let us define the truncated function as
\begin{equation}
    f_N (\lambda b) = \sum_{n = -N}^{+N} a_n e^{i n \lambda b} \quad.
\end{equation}
{\prop When evaluated on gaussian states \eqref{eq:eff_02} one has that
\begin{align}
    \left| \braket{f(\lambda b)} - \braket{f_N(\lambda b)} \right| \le 2 f_{\rm max}\ e^{-\sigma_b^2 \lambda^2 N^2} (1+\sO(e^{-\sigma_b^2\lambda^2}))\quad.
\end{align}
Therefore, the expectation value of the full series and the truncation agree extremely quickly as $N$ grows. We can  say the the difference between the two will be negligible as long as
\begin{equation}
   \boxed{ \lambda^2 \sigma^2_b N^2> 1}\quad.
\end{equation}
}
\proof It follows from Proposition \ref{jijo} that
\begin{align}
    \left| \braket{f(\lambda b)} - \braket{f_N(\lambda b)} \right| &= \left| \sum_{|n| > N} a_n e^{i n \lambda b_0} e^{-\sigma_b^2 \lambda^2 n^2} \right| \n\\
\n    &\le 2 \sum_{n=N+1}^{+\infty} |a_n|\ e^{-\sigma_b^2 \lambda^2 n^2} = 2\ e^{-\sigma_b^2 \lambda^2 N^2} \sum_{n=N+1}^{+\infty} |a_n|\ e^{-\sigma_b^2 \lambda^2 (n^2 - N^2)} \\
   \n &\le 2\ e^{-\sigma_b^2 \lambda^2 N^2} \sum_{n=N+1}^{+\infty} |a_n|\ e^{-\sigma_b^2 \lambda^2 (n-N)^2} \\
    \n&= 2\ e^{-\sigma_b^2 \lambda^2 N^2} \sum_{m=1}^{+\infty} |a_{m+N}|\ e^{-\sigma_b^2 \lambda^2 m^2} \\
    &\le 2 f_{\rm max}\ e^{-\sigma_b^2 \lambda^2 N^2} (1+\sO(e^{-\sigma_b^2\lambda^2}))\quad \square.
\end{align}

%
%

{\coloro \label{four}For a given function $f(\lambda b)$, and Gaussian states \eqref{eq:eff_02} we have that
\be
\braket{f_N(\lambda b)}\approx f(\lambda b_0) 
\ee  as long as 
\be
\boxed{\frac{1}{N^2}<\lambda^2 \sigma_b^2<\left|\frac {4f(\lambda b_0)}{f''(\lambda b_0)} \right|} 
\ee
which can be achieved for sufficiently large $N$.}
We therefore arrive at the conclusion that for any arbitrary function $f(\lambda b)$ we can find $N$ and $\sigma_p$ such that the Gaussian expectation value would agree with the desired accuracy with the function of our choice satisfying the minimal requirement \eqref{only-condition}.

{\prop\label{remyy} For arbitrary operators $O(\beta b)$ following time evolution rule holds
\ba
\frac{d\braket{O(\beta b)}}{ds}&\equiv&-i\braket{[O(\beta b), {\cal H}]}\n \\
&=&\frac{V_0}{4 \pi G \gamma }\frac{\braket{p_\phi^2}}{\pi G \gamma v_0^3} \frac{d\braket{O(\beta b)}}{db_0}+\sO\left(\frac{\beta^2}{v_0^4}\right)
\ea
where one needs to assume that $v_0\gg \sigma_b$ and  $v_0\gg N$.
}

%



\end{appendix}

\providecommand{\href}[2]{#2}\begingroup\raggedright\endgroup

\end{document}